\documentclass[aps,prl,twocolumn,superscriptaddress,amsfonts,amssymb,amsmath,showpacs,floatfix,nobibnotes,longbibliography,preprintnumbers]{revtex4-2}
\usepackage[utf8]{inputenc}
\usepackage[english]{babel}
\usepackage[T1]{fontenc}
\usepackage{natbib}
\usepackage[resetlabels]{multibib}
\usepackage[colorlinks=true,linkcolor=blue,citecolor=blue,urlcolor=blue]{hyperref}
\usepackage{graphicx}
\usepackage{epsfig}
\usepackage{color}
\usepackage{float}
\usepackage{mathtools}
\usepackage{bm,bbm}
\usepackage{isotope}
\usepackage[capitalize]{cleveref}

\DeclareUnicodeCharacter{202F}{\,}

\newcommand{\eq}[1]{Eq.~(\ref{#1})}

\begin{document}
\preprint{LA-UR-21-31426}

\title{Trends of Neutron Skins and Radii of Mirror Nuclei from First Principles}

\author{S.~J.~Novario}
\affiliation{Theoretical Division, Los Alamos National Laboratory,
  Los Alamos, New Mexico 87545, USA}

\author{D.~Lonardoni}
\thanks{Present affiliation: XCP-2: Eulerian Codes Group, Los Alamos National Laboratory,
  Los Alamos, New Mexico 87545, USA}
\affiliation{Theoretical Division, Los Alamos National Laboratory,
  Los Alamos, New Mexico 87545, USA}

\author{S.~Gandolfi}
\affiliation{Theoretical Division, Los Alamos National Laboratory,
  Los Alamos, New Mexico 87545, USA}

\author{G.~Hagen}
\affiliation{Physics Division, Oak Ridge National Laboratory,
  Oak Ridge, Tennessee 37831, USA} 
\affiliation{Department of Physics and Astronomy, University of Tennessee,
  Knoxville, Tennessee 37996, USA}

\begin{abstract}
  The neutron skin of atomic nuclei impacts the structure of
  neutron-rich nuclei, the equation of state of nucleonic matter, and
  the size of neutron stars. Here we predict the neutron skin of
  selected light- and medium-mass nuclei using coupled-cluster theory
  and the auxiliary field diffusion Monte Carlo method with two- and
  three-nucleon forces from chiral effective field theory. We find a
  linear correlation between the neutron skin and the isospin
  asymmetry in agreement with the liquid-drop model and compare with
  data. We also extract the linear relationship that describes the
  difference between neutron and proton radii of mirror nuclei and
  quantify the effect of charge symmetry breaking terms in the nuclear
  Hamiltonian. Our results for the mirror-difference charge radii and
  binding energies per nucleon agree with existing data.
\end{abstract}

\maketitle 

{\it Introduction.--- } The size of neutron stars and the
distributions of excess neutrons in medium-mass and heavy nuclei can
be linked to the microscopic forces between the constituent nucleons
that build up the atomic nucleus and the equation-of-state of
nucleonic matter~\cite{brown2000, horowitz2001a, horowitz2001b,
  Gandolfi:2012}. With recent advancements in both observational
astrophysics and experimental nuclear physics, investigating this link
and constraining nuclear models are now becoming possible. The advent
of multimessenger astronomy, established with the simultaneous
observation of a binary neutron star merger by the LIGO
gravitational-wave observatory~\cite{abbott2017a, abbott2017b} and
gamma-ray astronomy, has opened up the possibility of examining the
structure of neutron stars in new
detail~\cite{Almamun:2021}. Furthermore, experiments have been
designed to determine the neutron distributions in nuclei, including
the recent extractions of the neutron skin from measurements of the
parity-violating asymmetry in \isotope[208]{Pb}~\cite{abrahamyan2012,
  adhikari2021} and \isotope[48]{Ca}~\cite{adhikari2022}.

The neutron skin thickness $\Delta R_{\mathrm{np}}$ is defined as the
difference between the root-mean-squared (rms) point radii of the
neutron and proton density distributions. Collectively, the neutron
skin results from a balance between the inward pressure of the surface
tension on excess neutrons on the edge of the nucleus and outward
degeneracy pressure from excess neutrons within the core of the
nucleus. The same balance is reached in neutron stars with the inward
pressure of gravity. The relationship between this pressure and the
neutron density is quantified in the slope of the symmetry energy at
saturation density $L$. Therefore, this bulk property of neutron stars
should be related to the neutron skin of nuclei~\cite{brown2000,
  horowitz2001a, tsang2012, hagen2015, brown2017, fattoyev2018,
  bertulani2019}.

\begin{figure}[b]
  \setlength\abovecaptionskip{-0.2\baselineskip}
  \hspace*{-8pt}
  \includegraphics[width=0.5\textwidth]{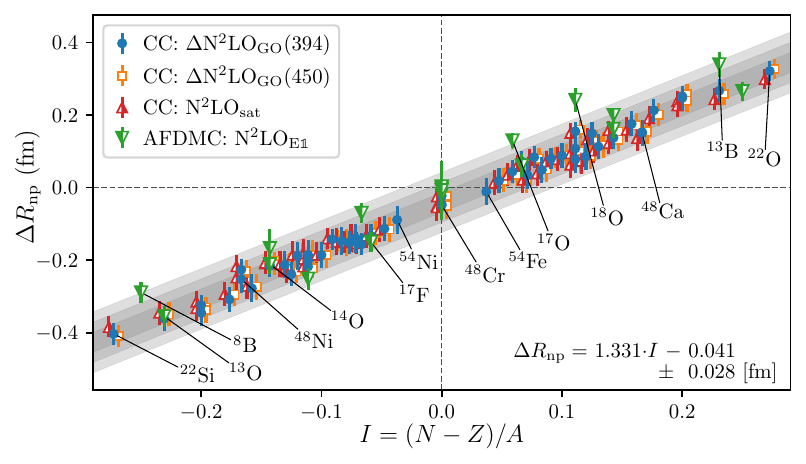}
  \caption[]{Neutron skin thickness plotted against the isospin
    asymmetry from \textit{ab initio} calculations. Coupled-cluster
    results of nuclei with $14\le A\le 56$ using the
    $\Delta$N$^2$LO$_{\rm GO}$(394), $\Delta$N$^2$LO$_{\rm GO}$(450),
    and N$^2$LO$_{\rm sat}$(450) interactions are shown as solid blue
    circles, empty orange squares, and right-filled red triangles,
    respectively. Left-filled green triangles represent auxiliary
    field diffusion Monte Carlo results for $6\le A\le 18$ using the
    N$^2$LO$_{E\mathbbm1}$ local chiral interaction. The linear
    regression is printed in the bottom right with $1\sigma$
    uncertainty, and the $1\sigma, 2\sigma$, and $3\sigma$ confidence
    levels are shown as gray bands. Select nuclei are labeled.}
  \label{skin_vs_asym}
\end{figure}

Recent high-precision measurements of charge radii in isotope chains
of potassium~\cite{koszorus2021}, calcium~\cite{garciaruiz2016,
  miller2019}, nickel ~\cite{malbrunot2022}, copper~\cite{bissell2016,
  groote2020}, and silver~\cite{reponen2021} have revealed that they
carry information about changes in shell structure, deformation, and
pairing effects. Charge radii can be easily probed with the well-known
electromagnetic interaction~\cite{dejager1974,
  angeli2013}. Conversely, measuring the neutron (weak-charge) radius
is more difficult, but can be accomplished using either strong or
electroweak probes, each with their own shortcomings. Hadronic probes
suffer from model-dependent uncertainties~\cite{piekarewicz2006}, and
electroweak probes, like those used in PREX~\cite{abrahamyan2012,
  adhikari2021} and CREX~\cite{adhikari2022}, measure the
parity-violating asymmetry at one momentum transfer and extract the
neutron skin using different models. As argued in
Ref.~\cite{reinhard2022} a clean comparison between theory and
experiment would be to compute this quantity instead of the neutron
skin. The issue with model dependencies in extracting the neutron skin
can be overcome in some cases~\cite{abrahamyan2012, adhikari2021}, but
precision measurements of the neutron radius for short-lived isotopes
remain elusive.

In this Letter, we study trends in the neutron skin thickness of
light- and medium-mass mirror nuclei with mass number $6\le A\le 56$
(shown in \cref{skin_vs_asym}) by computing the proton and neutron
radii using the \textit{ab initio} coupled cluster
(CC)~\cite{coester1958, coester1960, cizek1966, cizek1969,
  kuemmel1978, bartlett2007, hagen2013c} and auxiliary field diffusion
Monte Carlo (AFDMC) methods~\cite{schmidt1999, carlson2015,
  lonardoni2018a}, each with different nuclear interactions derived
from chiral effective field theory (described below). These methods
are validated by comparing our computed mirror-difference charge radii
and binding energies per nucleon with available data (shown in
\cref{Coul}). We then confirm the linear relation between neutron skin
thickness and isospin asymmetry from the liquid-drop model (LDM) and
compare this relation with the available experimental
data. Additionally, we also confirm a similar linear relation with the
difference between the neutron and proton radii of mirror partners and
connect this relation to the LDM. Finally, we estimate the
contribution of isospin-symmetry-breaking terms in the interaction by
calculating ground-state energies and radii with and without the
Coulomb interaction.

{\it Methods.--- } We describe nuclei as a collection of $A$ pointlike
nucleons of average mass $m$ interacting according to the
nonrelativistic intrinsic Hamiltonian
\begin{align}
  \label{hamiltonian}
  H = -\frac{\hbar^2}{2mA}\sum_{i<j} (\nabla_i - \nabla_j)^2+\sum_{i<j}V_{ij}+\sum_{i<j<k}V_{ijk}.
\end{align}
Here, $V_{ij}$ and $V_{ijk}$ are the nucleon-nucleon and three-nucleon
potentials, respectively, the former of which also includes the
Coulomb force. Interactions used in this Letter are derived from
chiral effective field theory (EFT)~\cite{epelbaum2009, machleidt2011,
  hebeler2011, ekstrom2013, entem2015, epelbaum2015, huther2020,
  soma2020}. As this work focuses on bulk properties of nuclei, we
choose a subset of chiral interactions that have been shown to
accurately reproduce these quantities. The first, N$^2$LO$_{\rm
  sat}$~\cite{ekstrom2015}, was optimized to few-body observables as
well as radii and ground-state energies of selected carbon and oxygen
isotopes. The second and third, $\Delta$N$^2$LO$_{\rm GO}(394)$ and
$\Delta$N$^2$LO$_{\rm GO}(450)$~\cite{payne2019, bagchi2020,
  jiang2020}, include explicit $\Delta$ isobars and were optimized to
few-body observables and properties of nucleonic matter, and use
momentum cutoff of $\Lambda=394$~MeV/$c$ and $\Lambda=450$~MeV/$c$,
respectively.

After establishing the nuclear interactions, we perform a Hartree-Fock
diagonalization from a spherical harmonic-oscillator basis and then
construct a prolate, axially-deformed product state built from natural
orbitals~\cite{tichai2019, novario2020}. We then normal-order the
Hamiltonian with respect to the reference state, denoted by
$|\Phi_0\rangle$, and retain one- and two-body terms~\cite{hagen2007a,
  roth2012}. Next, we use the coupled-cluster method in the
singles-and-doubles (CCSD) approximation to construct the
similarity-transformed Hamiltonian, $\overline{H}_N =
e^{-\hat{T}}{H_N}e^{\hat{T}}$, which decouples the reference state
from excitations around it. In the CCSD approximation the cluster
operator $\hat{T}$ is truncated at the $2p$--$2h$ excitation level.
Because $\hat{T}$ is asymmetric, $\overline{H}_N$ is non-Hermitian,
and the left ground state must be decoupled separately using $\langle
\Phi_0 \vert ( 1 + \hat{\Lambda} )$, where $\hat{\Lambda}$ is a
de-excitation analog to $\hat{T}$~\cite{bartlett2007, hagen2013c} and
is also truncated at the $2p$--$2h$ level. In this work we are
interested in computing the ground-state expectation value of the
squared intrinsic point-nucleon radius operator, $\hat{r}^2$. In CC
theory this operator is similarity transformed according to
$\overline{r}^{2}\equiv e^{-\hat{T}}\hat{r}^{2}e^{\hat{T}}$, and
applied between the left and right CC ground states, i.e., $\langle
\hat{r}^{2}\rangle \equiv
\langle\Phi_0|(1+\hat{\Lambda})\overline{r}^{2}|\Phi_0\rangle .$

We also perform AFDMC calculations of the point-proton and neutron
radii (see Ref.~\cite{lonardoni2018a} for more details) using a local
chiral nucleon-nucleon and three-nucleon interaction at
next-to-next-to-leading order, N$^2$LO$_{\rm
  E\mathbbm1}$~\cite{gezerlis2013, gezerlis2014, lynn2016,
  lonardoni2018a}. This interaction has been used to study the
ground-state properties of nuclei up to
\isotope[16]{O}~\cite{lynn2014, lynn2016, lynn2017, lonardoni2018,
  lonardoni2018a, lonardoni2018b, lynn2019a, lim2019, cruz2019,
  cruz2021}, few neutron systems~\cite{gandolfi2016, klos2016,
  gandolfi2017}, and neutron star matter~\cite{gezerlis2014, tews2016,
  lynn2016, buraczynski2016, buraczynski2017, riz2018, tews2018,
  tews2018a, gandolfi2018, tews2019, Lonardoni:2020,
  Riz2020}. Additionally, among other \textit{ab initio} methods~(see
Ref.~\cite{hergert2020} for a recent review), both CC and AFDMC have
been used to accurately calculate nuclear radii, e.g.,
\cite{koszorus2021, malbrunot2022} and \cite{lonardoni2018,
  gandolfi2020}, respectively.

{\it Results.--- } Our results for the neutron skin thickness of
selected nuclei with $6\le A\le 56$ are shown in
\cref{skin_vs_asym}. These data are plotted against the isospin
asymmetry, $I = (N-Z)/A$, with $N$ and $Z$ being the number of
neutrons and protons, respectively, and a linear relationship is
extracted. This relationship was first derived from the LDM in
Refs.~\cite{Myers:1969, Myers:1974, Myers:1977, Myers:1980} and
demonstrated with mean-field methods in Ref.~\cite{pethick1996}. We
fit the data using linear regression which is shown in
\cref{skin_vs_asym} as shaded gray bands at $1\sigma$, $2\sigma$, and
$3\sigma$ confidence levels. The fit uncertainties increase away from
$I = 0$, and the quoted uncertainties are taken from the maximum value
from within the data.

There are several distinct sources of uncertainties in the CC
calculations. First, the error associated with the finite size of the
employed model space is estimated by taking the difference between the
squared radii calculated in the model spaces of $N_{\rm max}=10,12$,
respectively, where $N_{\rm max} = (2n + l)_{\rm max}$ indicates the
maximum harmonic oscillator energy level. We also note that in this
work we utilize a basis with an oscillator energy of
$\hbar\omega=16\ \rm MeV$, which is suitable for all calculated
nuclei. The additional uncertainty associated with this choice is
estimated to be 1\%~\cite{kaufmann2020, novario2020}. Second, the
uncertainty associated with truncating the CC excitation operator at
the $2p$--$2h$ level amounts to another 1\%
uncertainty~\cite{novario2020, miorelli2018, kaufmann2020}. Finally,
the error from breaking rotational symmetry is estimated to be less
than 1\% on computed radii (see Ref.~\cite{hagen2022} for
details). Uncertainties on AFDMC results are statistical, and are
reported in this work at the $1\sigma$ confidence level. Errors on
radii also include an extrapolation uncertainty from mix estimates,
and those on binding energies from unconstrained evolution (see
Ref.~\cite{lonardoni2018a} for more details).

There are two main factors concerning the $A$-dependence of the linear
relation in \cref{skin_vs_asym}. First, it is expected that the
collective behavior which gives rise to the linear dependence on the
isospin asymmetry will be overshadowed by few-body effects for lighter
nuclei. We note that among the five light nuclei computed with both
methods, four neutron skin results were compatible. Second, there is
an $A$-dependence that accounts for the Coulomb repulsion of the
protons and can be derived from the LDM~\cite{Myers:1969,
  Myers:1980}. The first-order approximation to this effect is $\Delta
R^{\mathrm{Coul}}_{\mathrm{np}}\approx -Z A^{-1/3} \times (r_{0}c_{1}
/ 8Q^{*})$, where $r_{0}$ is the nuclear radius constant, $c_{1}$ is
the Coulomb energy coefficient, and $Q^{*}$ is the effective surface
stiffness coefficient (see Supplemental Material~\cite{supp} for
details). Using values from Ref.~\cite{Myers:1969}, this contribution
can be approximated as $\Delta R^{\mathrm{Coul}}_{\mathrm{np}} \approx
-Z A^{-1/3} \times 0.0033\ \mathrm{fm}$.

\begin{figure}[t]
  \setlength\abovecaptionskip{-0.2\baselineskip}
  \hspace*{-8pt}
  \includegraphics[width=0.5\textwidth]{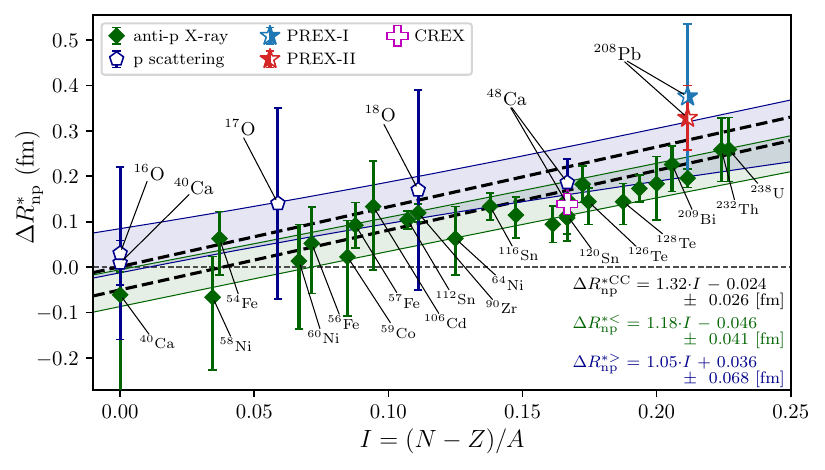}
  \caption[]{Coulomb-subtracted neutron skin thickness ($\Delta
    R^{*}_{\mathrm{np}} = \Delta R_{\mathrm{np}} + Z A^{-1/3} \times
    0.0033\ \mathrm{fm}$) plotted against the isospin asymmetry from
    available experimental data. Solid green diamonds are the results
    extracted from antiprotonic $X$-ray data~\cite{trzcinska2001,
      jastrzebbski2004} and open blue pentagons are from proton
    scattering~\cite{lapoux2016, zenihiro2018}. These data are
    accompanied by linear fits shown as green and blue bands,
    respectively, and both relations are printed in the bottom right
    ($\Delta R^{*<}_{\mathrm{np}}$ and $\Delta
    R^{*>}_{\mathrm{np}}$). The right-filled blue star and the
    left-filled red star are the values for \isotope[208]{Pb} from
    PREX-I~\cite{abrahamyan2012} and PREX-II~\cite{adhikari2021},
    respectively. The open cross is the value for \isotope[48]{Ca}
    from CREX~\cite{adhikari2022}. These experimental data are
    compared with our \textit{ab initio} results, shown as dashed
    black lines from the $1\sigma$ limits of a linear regression.}
  \label{skin_exp}
\end{figure}

To test the validity of our linear relationship between neutron skin
thickness and the isospin asymmetry, we plot the Coulomb-subtracted
neutron skin thickness ($\Delta R^{*}_{\mathrm{np}} = \Delta
R_{\mathrm{np}} - \Delta R^{\mathrm{Coul}}_{\mathrm{np}}$) in
\cref{skin_exp} for experimental data of neutron-rich nuclei along
with the linear regression of our data at the $1\sigma$ confidence
level, which is shown by two dashed black lines. Most of the
experimental data are extracted from x-ray spectroscopy following
antiprotonic annihilation~\cite{trzcinska2001,
  jastrzebbski2004}. These data are used to fit a linear relationship
which is shown as a green band and denoted as $\Delta
R^{*<}_{\mathrm{np}}$. Additional data from proton scattering are also
included~\cite{lapoux2016, zenihiro2018} and separately fit, which is
shown as a blue band and denoted as $\Delta
R^{*>}_{\mathrm{np}}$. Also shown are the results from
CREX~\cite{adhikari2022}, PREX-I~\cite{abrahamyan2012}, and
PREX-II~\cite{adhikari2021}, which extract the neutron radius of
$^{48}$Ca and \isotope[208]{Pb} from parity-violating electron
scattering, respectively.

Our linear fit overlaps significantly with both the antiprotonic x-ray
and proton-scattering data but has a larger slope. In the LDM, this
slope is related to the ratio of the symmetry-energy coefficient $J$
and the effective stiffness coefficient ($\approx 3r_{0}J/2Q*$). Using
the slope from our results, we obtain a ratio of $J / Q \approx 1.59$
which is larger than those obtained with many Skyrme
forces~\cite{Warda:2009} but consistent with the values from
Ref.~\cite{Myers:1969}. Using our linear relationship to predict the
neutron skin of \isotope[208]{Pb}, we obtain $\Delta
R_{\mathrm{np}}\left(^{208}\mathrm{Pb}\right) = 0.210 \pm
0.026\ \mathrm{fm}$, which is consistent with the recent \textit{ab
  initio} prediction in Ref.~\cite{hu2022}.

\begin{figure}[t]
  \setlength\abovecaptionskip{-0.2\baselineskip}
  \hspace*{-8pt}
  \includegraphics[width=0.5\textwidth]{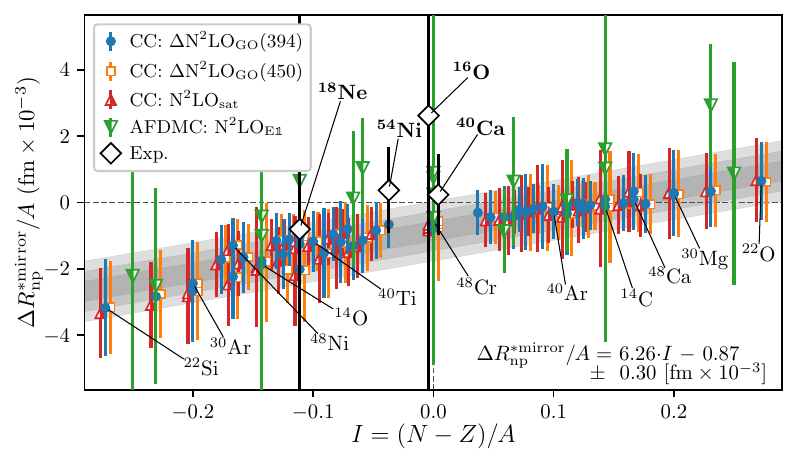}
  \caption[]{Coulomb-subtracted mirror neutron skin thickness $\Delta
    R^{*\mathrm{mirror}}_{\mathrm{np}}/A = (\Delta
    R^{\mathrm{mirror}}_{\mathrm{np}} + Z A^{-1/3} \times
    0.0033\ \mathrm{fm})/A$ plotted against the isospin asymmetry. The
    data markers are the same as those in \cref{skin_vs_asym}. The
    experimental data for $^{16}$O ($^{18}$Ne), $^{40}$Ca, and
    $^{54}$Ni are extracted from Refs.~\cite{lapoux2016,angeli2013},
    \cite{jastrzebbski2004,angeli2013}, and
    \cite{trzcinska2001,pineda2021}, respectively. The linear
    regression is printed in the bottom right with $1\sigma$
    uncertainty, and the $1\sigma, 2\sigma$, and $3\sigma$ confidence
    levels are shown as gray bands. Selected nuclei are labeled.}
  \label{rnmirror_vs_rp}
\end{figure}

The bulk properties that give rise to the neutron skin can also be
utilized to establish a relationship between the point-proton radius
of a nucleus, $R_{\mathrm{p}}$ and the point-neutron radius of its
mirror partner, $R^{\mathrm{mirror}}_{\mathrm{n}}$. In the limit of
conserved isospin symmetry, these two quantities should be exactly
equal so that $\Delta R^{\mathrm{mirror}}_{\mathrm{np}} =
R^{\mathrm{mirror}}_{\mathrm{n}} - R_{\mathrm{p}} = 0$. The
first-order correction to this limit from the
LDM~\cite{Myers:1969,Myers:1980} is the same as for the neutron skin
thickness, $\Delta R^{\mathrm{Coul}}_{\mathrm{np}}$. However, in this
case the Coulomb-subtracted difference, $\Delta
R^{*\mathrm{mirror}}_{\mathrm{np}}$ should be proportional to $I
\times A$ such that $\Delta R^{*\mathrm{mirror}}_{\mathrm{np}}/A
\approx I \times (r_{0}c_{1}/K)$, where $K$ is the compressibility
coefficient (see Supplemental Material~\cite{supp} for details).

To test this relationship, we plot the Coulomb-subtracted difference,
$\Delta R^{*\mathrm{mirror}}_{\mathrm{np}}$, against the isospin
asymmetry in \cref{rnmirror_vs_rp}. The nonzero intercept results from
higher-order Coulomb corrections and shell effects. The slope is
related to the compressibility coefficient, which can be extracted as
$K \approx 138\ \mathrm{MeV}$, a smaller value compared with the range
from the interactions used in this work ($150\ \mathrm{MeV} \le K \le
250\ \mathrm{MeV}$ ~\cite{jiang2020, hagen2015, Lonardoni:2020}). In
addition to our results, we also show experimental data for $^{16}$O,
$^{18}$Ne, $^{40}$Ca, and $^{54}$Ni. The proton radii are extracted
from Refs.~\cite{angeli2013} and \cite{pineda2021} for $^{54}$Ni. The
neutron radii are then extracted from the neutron skins in
Refs.~\cite{lapoux2016, jastrzebbski2004, trzcinska2001} and charge
radii in Ref.~\cite{angeli2013}. This relationship in
\cref{rnmirror_vs_rp} can be employed to determine the elusive neutron
radius of a nucleus by more easily measuring the proton radius of its
mirror nucleus. While some of the proton-rich nuclei in these
calculations are near the drip-line, the Coulomb barrier ensures that
their ground states are weakly coupled to nearby proton-continuum
states. Therefore, these nuclei can be accurately described by
bound-state methods including CC ~\cite{holt2013b, morris2018,
  Michel:2019, Randhawa:2019, Stroberg:2021}.

\begin{figure}[t]
  \setlength\abovecaptionskip{-0.2\baselineskip}
  \hspace*{-8pt}
  \includegraphics[width=0.5\textwidth]{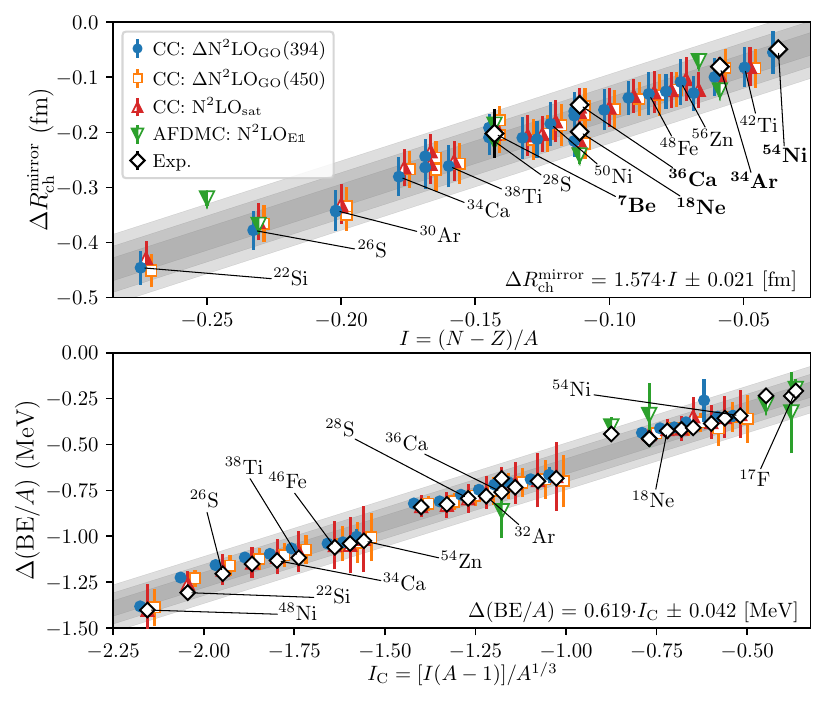}
  \caption[]{Top panel: Mirror-difference charge radii plotted against
    the isospin asymmetry compared with experiment. The data markers
    are the same as those in \cref{skin_vs_asym}. The linear
    regression of our data is printed in the bottom left with
    $1\sigma$ uncertainty, and the $1\sigma, 2\sigma$, and $3\sigma$
    confidence levels are shown as gray bands. Selected
    neutron-deficient nuclei are labeled. Experimental data are shown
    as open white diamonds, taken from Ref.~\cite{angeli2013} except
    for those of \isotope[36]{Ca}~\cite{miller2019} and
    \isotope[54]{Ni}~\cite{pineda2021}. Bottom panel:
    Mirror-difference binding energy per nucleon
    $\Delta(\mathrm{BE}/A)$ as a function of the Coulomb
    asymmetry. Experimental data are taken from Ref.~\cite{wang2017}.}
  \label{Coul}
\end{figure}

The near equivalence between $R^{\mathrm{mirror}}_{\mathrm{n}}$ and
$R_{\mathrm{p}}$, shown in \cref{rnmirror_vs_rp}, suggests a
correspondence between the neutron skin and the mirror-difference
charge radii, $\Delta
R_{\mathrm{ch}}^{\mathrm{mirror}}({}^{A}_{Z}\mathrm{X}_{N}) = \left[
  R_{\mathrm{ch}} \left( {}^{A}_{N}\mathrm{Y}_{Z} \right) -
  R_{\mathrm{ch}} \left( {}^{A}_{Z}\mathrm{X}_{N} \right) \right]$. We
compute the mirror-difference charge radii for selected nuclei and
compare with available experimental data, shown in~\cref{Coul}. The
charge radius is computed using the formula $R_{\mathrm{ch}}^2 =
R_{\mathrm{p}}^2 + \langle r_{\mathrm{p}}^2 \rangle + {N\over Z}
\langle r_{\mathrm{n}}^2 \rangle + \langle r_{\mathrm{DF}}^2 \rangle +
\langle r_{\mathrm{SO}}^2 \rangle$ where the spin-orbit correction,
$\langle r_{\mathrm{SO}}^2 \rangle$, is calculated with the CC
method~\cite{hagen2015}, and $\langle r_{\mathrm{p}}^2\rangle=
0.709$~fm$^2$, $\langle r_{\mathrm{n}}^2\rangle=-0.106$~fm$^2$, and
$\langle r_{\mathrm{DF}}^2\rangle=3/(4m^2)=0.033$~fm$^2$ are the
charge radius squared of the proton~\cite{pohl2010,xiong2019}, the
neutron~\cite{filin2020}, and the Darwin-Foldy term, respectively. Our
results agree well with the sparse data, most of which were taken from
Ref.~\cite{angeli2013} except for those of
\isotope[36]{Ca}~\cite{miller2019} and
\isotope[54]{Ni}~\cite{pineda2021}. Because $\Delta
R_{\mathrm{ch}}^{\mathrm{mirror}}$ is an analog of the neutron skin,
it also has a linear relation to the isospin asymmetry. Also shown in
\cref{Coul} is the comparison of our results with data from
Ref.~\cite{wang2017} of mirror-difference binding energies per nucleon
$\left\{ \Delta \left( \mathrm{BE}/A \right)\left(
{}^{A}_{Z}\mathrm{X}_{N} \right) = \left[\mathrm{BE}\left(
  {}^{A}_{Z}\mathrm{X}_{N} \right) - \mathrm{BE}\left(
  {}^{A}_{N}\mathrm{Y}_{Z}\right)\right]/A\right\}$. This quantity
exhibits a linear relationship with the Coulomb asymmetry,
$I_{\mathrm{C}} = [I (A-1)]/A^{1/3}$, which stems from the only
mirror-asymmetric term in the LDM. The absolute binding energies per
nucleon and charge radii are also compared to experimental data in the
Supplemental Material~\cite{supp}.

Finally, we also study the extent of the isospin-symmetry-breaking
terms in the nuclear interaction by computing ground-state energies
and radii of select mirror nuclei using CC theory with and without the
Coulomb term. We show these results for $A=42-48$ using
$\Delta$N$^2$LO$_{\rm GO}$(394) in \cref{tab:NoCoul48}, for the
neutron skin thickness and the mirror-difference binding energy per
nucleon; results for additional nuclei and $\Delta$N$^2$LO$_{\rm
  GO}$(450) are shown in the Supplemental Material~\cite{supp}. The
neutron skin thickness computed without the Coulomb term still
exhibits a linear relationship with the isospin asymmetry. However,
unlike the relation shown in \cref{Coul}, $\Delta(\mathrm{BE}/A)$
computed without the Coulomb term also exhibits a linear relationship
with the isospin asymmetry instead of $I_{\mathrm{C}}$. This quantity
entirely reflects the effects of the isospin-symmetry-breaking terms
in the chiral interaction, which account for $\sim 1\%$ of
$\Delta(\mathrm{BE}/A)$. These results might be useful to constrain
the charge-symmetry-breaking term in Skyrme density functional theory
according to Ref.~\cite{naito2022}.

\begin{table}[h]
  \caption{Neutron skin thickness ($\Delta R_{\mathrm{np}}$) and
    mirror-difference binding energy per nucleon ($\Delta ({\rm
      BE}/A)$) of select nuclei using $\Delta$N$^2$LO$_{\rm GO}(394)$
    with and without the Coulomb term (Coul.).}\label{tab:NoCoul48}
  \begin{ruledtabular}
    \begin{tabular}{c c c c c}
      & \multicolumn{2}{c}{$\underline{\Delta R_{\mathrm{np}}\ \rm (fm)}$} & \multicolumn{2}{c}{$\underline{\Delta ({\rm BE}/A)\ ({\rm MeV})}$} \\
      & w/ Coul. & w/o Coul. & w/ Coul. & w/o Coul. \\ 
      \hline & \\ [-2ex]
      $^{42}\rm Ca$ &  0.019(35)  &  0.060(34)  &  0.349(18)  &  0.004(15)  \\
      $^{46}\rm Ca$ &  0.113(35)  &  0.151(35)  &  1.039(25)  &  0.011(20)  \\
      $^{48}\rm Ca$ &  0.150(36)  &  0.186(35)  &  1.382(28)  &  0.015(23)  \\
      $^{48}\rm Ti$ &  0.049(37)  &  0.090(36)  &  0.689(29)  &  0.007(24)  \\
      $^{48}\rm Cr$ & -0.048(38)  & -0.001(36)  &  0.000(30)  &  0.000(26)  \\ 
    \end{tabular}
  \end{ruledtabular}
\end{table}

{\it Conclusions.---} We performed CC and AFDMC calculations of
neutron skins of a wide range of light- and medium-mass nuclei using
different interactions from chiral EFT, and found a robust linear
relationship with the isospin asymmetry in agreement with the
liquid-drop model. When extrapolated to heavy nuclei, this
relationship is consistent with the available data. Additionally, we
found a similar linear relationship between the neutron and proton
radii of mirror nuclei pairs which can be used to estimate neutron
radii of rare isotopes not yet accessible by experiment. Our
calculations are in good agreement with available data for the
difference in charge radii and binding energies of mirror
nuclei. Finally, we estimated the contributions from
charge-symmetry-breaking terms in the nuclear interaction which can be
useful for constraining those terms in Skyrme density functionals.

\begin{acknowledgments}
We would like to thank J.~Carlson, I.~Tews, and R.F.~Garcia Ruiz for
helpful discussions. The work of S.G., S.N., and D.L. was supported by
the DOE NUCLEI SciDAC Program, and by the DOE Early Career Research
Program.  The work of S.G. was also supported by U.S.~Department of
Energy, Office of Science, Office of Nuclear Physics, under Contract
No.~DE-AC52-06NA25396. G.H. was supported by the Office of Nuclear
Physics, U.S. Department of Energy, under Grant No. DE-SC0018223
(NUCLEI SciDAC-4 collaboration), and under contract DE-AC05-00OR22725
with UT-Battelle, LLC (Oak Ridge National Laboratory). Computer time
was provided by the Innovative and Novel Computational Impact on
Theory and Experiment (INCITE) program. This research used resources
of the Oak Ridge Leadership Computing Facility located at ORNL, which
is supported by the Office of Science of the Department of Energy
under Contract No. DE-AC05-00OR22725. This research also used
resources provided by the Los Alamos National Laboratory Institutional
Computing Program, which is supported by the U.S. Department of Energy
National Nuclear Security Administration under Contract
No. 89233218CNA000001.
\end{acknowledgments}


\begin{thebibliography}{105}%
\makeatletter
\providecommand \@ifxundefined [1]{%
 \@ifx{#1\undefined}
}%
\providecommand \@ifnum [1]{%
 \ifnum #1\expandafter \@firstoftwo
 \else \expandafter \@secondoftwo
 \fi
}%
\providecommand \@ifx [1]{%
 \ifx #1\expandafter \@firstoftwo
 \else \expandafter \@secondoftwo
 \fi
}%
\providecommand \natexlab [1]{#1}%
\providecommand \enquote  [1]{``#1''}%
\providecommand \bibnamefont  [1]{#1}%
\providecommand \bibfnamefont [1]{#1}%
\providecommand \citenamefont [1]{#1}%
\providecommand \href@noop [0]{\@secondoftwo}%
\providecommand \href [0]{\begingroup \@sanitize@url \@href}%
\providecommand \@href[1]{\@@startlink{#1}\@@href}%
\providecommand \@@href[1]{\endgroup#1\@@endlink}%
\providecommand \@sanitize@url [0]{\catcode `\\12\catcode `\$12\catcode
  `\&12\catcode `\#12\catcode `\^12\catcode `\_12\catcode `\%12\relax}%
\providecommand \@@startlink[1]{}%
\providecommand \@@endlink[0]{}%
\providecommand \url  [0]{\begingroup\@sanitize@url \@url }%
\providecommand \@url [1]{\endgroup\@href {#1}{\urlprefix }}%
\providecommand \urlprefix  [0]{URL }%
\providecommand \Eprint [0]{\href }%
\providecommand \doibase [0]{https://doi.org/}%
\providecommand \selectlanguage [0]{\@gobble}%
\providecommand \bibinfo  [0]{\@secondoftwo}%
\providecommand \bibfield  [0]{\@secondoftwo}%
\providecommand \translation [1]{[#1]}%
\providecommand \BibitemOpen [0]{}%
\providecommand \bibitemStop [0]{}%
\providecommand \bibitemNoStop [0]{.\EOS\space}%
\providecommand \EOS [0]{\spacefactor3000\relax}%
\providecommand \BibitemShut  [1]{\csname bibitem#1\endcsname}%
\let\auto@bib@innerbib\@empty
\bibitem [{\citenamefont {Alex~Brown}(2000)}]{brown2000}%
  \BibitemOpen
  \bibfield  {author} {\bibinfo {author} {\bibfnamefont {B.~A.}~\bibnamefont
  {Brown}},\ }\href {https://doi.org/10.1103/PhysRevLett.85.5296}
  {\bibfield  {journal} {\bibinfo  {journal} {Phys. Rev. Lett.}\ }\textbf
  {\bibinfo {volume} {85}},\ \bibinfo {pages} {5296} (\bibinfo {year}
  {2000})}\BibitemShut {NoStop}%
\bibitem [{\citenamefont {Horowitz}\ and\ \citenamefont
  {Piekarewicz}(2001{\natexlab{a}})}]{horowitz2001a}%
  \BibitemOpen
  \bibfield  {author} {\bibinfo {author} {\bibfnamefont {C.~J.}\ \bibnamefont
  {Horowitz}}\ and\ \bibinfo {author} {\bibfnamefont {J.}~\bibnamefont
  {Piekarewicz}},\ }\href {https://doi.org/10.1103/PhysRevLett.86.5647}
  {\bibfield  {journal} {\bibinfo  {journal} {Phys. Rev. Lett.}\ }\textbf
  {\bibinfo {volume} {86}},\ \bibinfo {pages} {5647} (\bibinfo {year}
  {2001}{\natexlab{a}})}\BibitemShut {NoStop}%
\bibitem [{\citenamefont {Horowitz}\ and\ \citenamefont
  {Piekarewicz}(2001{\natexlab{b}})}]{horowitz2001b}%
  \BibitemOpen
  \bibfield  {author} {\bibinfo {author} {\bibfnamefont {C.~J.}\ \bibnamefont
  {Horowitz}}\ and\ \bibinfo {author} {\bibfnamefont {J.}~\bibnamefont
  {Piekarewicz}},\ }\href {https://doi.org/10.1103/PhysRevC.64.062802}
  {\bibfield  {journal} {\bibinfo  {journal} {Phys. Rev. C}\ }\textbf {\bibinfo
  {volume} {64}},\ \bibinfo {pages} {062802(R)} (\bibinfo {year}
  {2001}{\natexlab{b}})}\BibitemShut {NoStop}%
\bibitem [{\citenamefont {Gandolfi}\ \emph {et~al.}(2012)\citenamefont
  {Gandolfi}, \citenamefont {Carlson},\ and\ \citenamefont
  {Reddy}}]{Gandolfi:2012}%
  \BibitemOpen
  \bibfield  {author} {\bibinfo {author} {\bibfnamefont {S.}~\bibnamefont
  {Gandolfi}}, \bibinfo {author} {\bibfnamefont {J.}~\bibnamefont {Carlson}},\
  and\ \bibinfo {author} {\bibfnamefont {S.}~\bibnamefont {Reddy}},\ }\href
  {https://doi.org/10.1103/PhysRevC.85.032801} {\bibfield  {journal} {\bibinfo
  {journal} {Phys. Rev. C}\ }\textbf {\bibinfo {volume} {85}},\ \bibinfo
  {pages} {032801(R)} (\bibinfo {year} {2012})}\BibitemShut {NoStop}%
\bibitem [{\citenamefont {Abbott}\ \emph
  {et~al.}(2017{\natexlab{a}})\citenamefont {Abbott}, \citenamefont {Abbott},
  \citenamefont {Abbott}, \citenamefont {Acernese}, \citenamefont {Ackley},
  \citenamefont {Adams}, \citenamefont {Adams}, \citenamefont {Addesso},
  \citenamefont {Adhikari}, \citenamefont {Adya}, \citenamefont {Affeldt},
  \citenamefont {Afrough}, \citenamefont {Agarwal}, \citenamefont {Agathos},
  \citenamefont {Agatsuma}\ \emph {et~al.}}]{abbott2017a}%
  \BibitemOpen
  \bibfield  {author} {\bibinfo {author} {\bibfnamefont {B.~P.}\ \bibnamefont
  {Abbott}}, \bibinfo {author} {\bibfnamefont {R.}~\bibnamefont {Abbott}},
  \bibinfo {author} {\bibfnamefont {T.~D.}\ \bibnamefont {Abbott}}, \bibinfo
  {author} {\bibfnamefont {F.}~\bibnamefont {Acernese}}, \bibinfo {author}
  {\bibfnamefont {K.}~\bibnamefont {Ackley}}, \bibinfo {author} {\bibfnamefont
  {C.}~\bibnamefont {Adams}}, \bibinfo {author} {\bibfnamefont
  {T.}~\bibnamefont {Adams}}, \bibinfo {author} {\bibfnamefont
  {P.}~\bibnamefont {Addesso}}, \bibinfo {author} {\bibfnamefont {R.~X.}\
  \bibnamefont {Adhikari}}, \bibinfo {author} {\bibfnamefont {V.~B.}\
  \bibnamefont {Adya}}, \bibinfo {author} {\bibfnamefont {C.}~\bibnamefont
  {Affeldt}}, \bibinfo {author} {\bibfnamefont {M.}~\bibnamefont {Afrough}},
  \bibinfo {author} {\bibfnamefont {B.}~\bibnamefont {Agarwal}}, \bibinfo
  {author} {\bibfnamefont {M.}~\bibnamefont {Agathos}}, \bibinfo {author}
  {\bibfnamefont {K.}~\bibnamefont {Agatsuma}}\ \emph {et~al.} (\bibinfo
  {collaboration} {LIGO Scientific Collaboration and Virgo Collaboration}),\
  }\href {https://doi.org/10.1103/PhysRevLett.119.161101} {\bibfield  {journal}
  {\bibinfo  {journal} {Phys. Rev. Lett.}\ }\textbf {\bibinfo {volume} {119}},\
  \bibinfo {pages} {161101} (\bibinfo {year} {2017}{\natexlab{a}})}\BibitemShut
  {NoStop}%
\bibitem [{\citenamefont {Abbott}\ \emph
  {et~al.}(2017{\natexlab{b}})\citenamefont {Abbott}, \citenamefont {Abbott},
  \citenamefont {Abbott}, \citenamefont {Acernese}, \citenamefont {Ackley},
  \citenamefont {Adams}, \citenamefont {Adams}, \citenamefont {Addesso},
  \citenamefont {Adhikari}, \citenamefont {Adya}, \citenamefont {Affeldt},
  \citenamefont {Afrough}, \citenamefont {Agarwal}, \citenamefont {Agathos},
  \citenamefont {Agatsuma}\ \emph {et~al.}}]{abbott2017b}%
  \BibitemOpen
  \bibfield  {author} {\bibinfo {author} {\bibfnamefont {B.~P.}\ \bibnamefont
  {Abbott}}, \bibinfo {author} {\bibfnamefont {R.}~\bibnamefont {Abbott}},
  \bibinfo {author} {\bibfnamefont {T.~D.}\ \bibnamefont {Abbott}}, \bibinfo
  {author} {\bibfnamefont {F.}~\bibnamefont {Acernese}}, \bibinfo {author}
  {\bibfnamefont {K.}~\bibnamefont {Ackley}}, \bibinfo {author} {\bibfnamefont
  {C.}~\bibnamefont {Adams}}, \bibinfo {author} {\bibfnamefont
  {T.}~\bibnamefont {Adams}}, \bibinfo {author} {\bibfnamefont
  {P.}~\bibnamefont {Addesso}}, \bibinfo {author} {\bibfnamefont {R.~X.}\
  \bibnamefont {Adhikari}}, \bibinfo {author} {\bibfnamefont {V.~B.}\
  \bibnamefont {Adya}}, \bibinfo {author} {\bibfnamefont {C.}~\bibnamefont
  {Affeldt}}, \bibinfo {author} {\bibfnamefont {M.}~\bibnamefont {Afrough}},
  \bibinfo {author} {\bibfnamefont {B.}~\bibnamefont {Agarwal}}, \bibinfo
  {author} {\bibfnamefont {M.}~\bibnamefont {Agathos}}, \bibinfo {author}
  {\bibfnamefont {K.}~\bibnamefont {Agatsuma}}\ \emph {et~al.},\ }\href
  {https://doi.org/10.3847/2041-8213/aa91c9} {\bibfield  {journal} {\bibinfo
  {journal} {Astrophys. J.}\ }\textbf {\bibinfo {volume} {848}},\ \bibinfo {pages} {L12}
  (\bibinfo {year} {2017}{\natexlab{b}})}\BibitemShut {NoStop}%
\bibitem [{\citenamefont {Al-Mamun}\ \emph {et~al.}(2021)\citenamefont
  {Al-Mamun}, \citenamefont {Steiner}, \citenamefont {N\"attil\"a},
  \citenamefont {Lange}, \citenamefont {O'Shaughnessy}, \citenamefont {Tews},
  \citenamefont {Gandolfi}, \citenamefont {Heinke},\ and\ \citenamefont
  {Han}}]{Almamun:2021}%
  \BibitemOpen
  \bibfield  {author} {\bibinfo {author} {\bibfnamefont {M.}~\bibnamefont
  {Al-Mamun}}, \bibinfo {author} {\bibfnamefont {A.~W.}\ \bibnamefont
  {Steiner}}, \bibinfo {author} {\bibfnamefont {J.}~\bibnamefont
  {N\"attil\"a}}, \bibinfo {author} {\bibfnamefont {J.}~\bibnamefont {Lange}},
  \bibinfo {author} {\bibfnamefont {R.}~\bibnamefont {O'Shaughnessy}}, \bibinfo
  {author} {\bibfnamefont {I.}~\bibnamefont {Tews}}, \bibinfo {author}
  {\bibfnamefont {S.}~\bibnamefont {Gandolfi}}, \bibinfo {author}
  {\bibfnamefont {C.}~\bibnamefont {Heinke}},\ and\ \bibinfo {author}
  {\bibfnamefont {S.}~\bibnamefont {Han}},\ }\href
  {https://doi.org/10.1103/PhysRevLett.126.061101} {\bibfield  {journal}
  {\bibinfo  {journal} {Phys. Rev. Lett.}\ }\textbf {\bibinfo {volume} {126}},\
  \bibinfo {pages} {061101} (\bibinfo {year} {2021})}\BibitemShut {NoStop}%
\bibitem [{\citenamefont {Abrahamyan}\ \emph {et~al.}(2012)\citenamefont
  {Abrahamyan}, \citenamefont {Ahmed}, \citenamefont {Albataineh},
  \citenamefont {Aniol}, \citenamefont {Armstrong}, \citenamefont {Armstrong},
  \citenamefont {Averett}, \citenamefont {Babineau}, \citenamefont {Barbieri},
  \citenamefont {Bellini}, \citenamefont {Beminiwattha}, \citenamefont
  {Benesch}, \citenamefont {Benmokhtar}, \citenamefont {Bielarski},
  \citenamefont {Boeglin}\ \emph {et~al.}}]{abrahamyan2012}%
  \BibitemOpen
  \bibfield  {author} {\bibinfo {author} {\bibfnamefont {S.}~\bibnamefont
  {Abrahamyan}}, \bibinfo {author} {\bibfnamefont {Z.}~\bibnamefont {Ahmed}},
  \bibinfo {author} {\bibfnamefont {H.}~\bibnamefont {Albataineh}}, \bibinfo
  {author} {\bibfnamefont {K.}~\bibnamefont {Aniol}}, \bibinfo {author}
  {\bibfnamefont {D.~S.}\ \bibnamefont {Armstrong}}, \bibinfo {author}
  {\bibfnamefont {W.}~\bibnamefont {Armstrong}}, \bibinfo {author}
  {\bibfnamefont {T.}~\bibnamefont {Averett}}, \bibinfo {author} {\bibfnamefont
  {B.}~\bibnamefont {Babineau}}, \bibinfo {author} {\bibfnamefont
  {A.}~\bibnamefont {Barbieri}}, \bibinfo {author} {\bibfnamefont
  {V.}~\bibnamefont {Bellini}}, \bibinfo {author} {\bibfnamefont
  {R.}~\bibnamefont {Beminiwattha}}, \bibinfo {author} {\bibfnamefont
  {J.}~\bibnamefont {Benesch}}, \bibinfo {author} {\bibfnamefont
  {F.}~\bibnamefont {Benmokhtar}}, \bibinfo {author} {\bibfnamefont
  {T.}~\bibnamefont {Bielarski}}, \bibinfo {author} {\bibfnamefont
  {W.}~\bibnamefont {Boeglin}}\ \emph {et~al.} (\bibinfo {collaboration} {PREX
  Collaboration}),\ }\href {https://doi.org/10.1103/PhysRevLett.108.112502}
  {\bibfield  {journal} {\bibinfo  {journal} {Phys. Rev. Lett.}\ }\textbf
  {\bibinfo {volume} {108}},\ \bibinfo {pages} {112502} (\bibinfo {year}
  {2012})}\BibitemShut {NoStop}%
\bibitem [{\citenamefont {Adhikari}\ \emph {et~al.}(2021)\citenamefont
  {Adhikari}, \citenamefont {Albataineh}, \citenamefont {Androic},
  \citenamefont {Aniol}, \citenamefont {Armstrong}, \citenamefont {Averett},
  \citenamefont {Ayerbe~Gayoso}, \citenamefont {Barcus}, \citenamefont
  {Bellini}, \citenamefont {Beminiwattha}, \citenamefont {Benesch},
  \citenamefont {Bhatt}, \citenamefont {Bhatta~Pathak}, \citenamefont
  {Bhetuwal}, \citenamefont {Blaikie}\ \emph {et~al.}}]{adhikari2021}%
  \BibitemOpen
  \bibfield  {author} {\bibinfo {author} {\bibfnamefont {D.}~\bibnamefont
  {Adhikari}}, \bibinfo {author} {\bibfnamefont {H.}~\bibnamefont
  {Albataineh}}, \bibinfo {author} {\bibfnamefont {D.}~\bibnamefont {Androic}},
  \bibinfo {author} {\bibfnamefont {K.}~\bibnamefont {Aniol}}, \bibinfo
  {author} {\bibfnamefont {D.~S.}\ \bibnamefont {Armstrong}}, \bibinfo {author}
  {\bibfnamefont {T.}~\bibnamefont {Averett}}, \bibinfo {author} {\bibfnamefont
  {C.}~\bibnamefont {Ayerbe~Gayoso}}, \bibinfo {author} {\bibfnamefont
  {S.}~\bibnamefont {Barcus}}, \bibinfo {author} {\bibfnamefont
  {V.}~\bibnamefont {Bellini}}, \bibinfo {author} {\bibfnamefont {R.~S.}\
  \bibnamefont {Beminiwattha}}, \bibinfo {author} {\bibfnamefont {J.~F.}\
  \bibnamefont {Benesch}}, \bibinfo {author} {\bibfnamefont {H.}~\bibnamefont
  {Bhatt}}, \bibinfo {author} {\bibfnamefont {D.}~\bibnamefont
  {Bhatta~Pathak}}, \bibinfo {author} {\bibfnamefont {D.}~\bibnamefont
  {Bhetuwal}}, \bibinfo {author} {\bibfnamefont {B.}~\bibnamefont {Blaikie}}\
  \emph {et~al.} (\bibinfo {collaboration} {PREX Collaboration}),\ }\href
  {https://doi.org/10.1103/PhysRevLett.126.172502} {\bibfield  {journal}
  {\bibinfo  {journal} {Phys. Rev. Lett.}\ }\textbf {\bibinfo {volume} {126}},\
  \bibinfo {pages} {172502} (\bibinfo {year} {2021})}\BibitemShut {NoStop}%
\bibitem [{\citenamefont {Adhikari}\ \emph {et~al.}(2022)\citenamefont
  {Adhikari}, \citenamefont {Albataineh}, \citenamefont {Androic},
  \citenamefont {Aniol}, \citenamefont {Armstrong}, \citenamefont {Averett},
  \citenamefont {Ayerbe~Gayoso}, \citenamefont {Barcus}, \citenamefont
  {Bellini}, \citenamefont {Beminiwattha}, \citenamefont {Benesch},
  \citenamefont {Bhatt}, \citenamefont {Bhatta~Pathak}, \citenamefont
  {Bhetuwal}, \citenamefont {Blaikie}\ \emph {et~al.}}]{adhikari2022}%
  \BibitemOpen
  \bibfield  {author} {\bibinfo {author} {\bibfnamefont {D.}~\bibnamefont
  {Adhikari}}, \bibinfo {author} {\bibfnamefont {H.}~\bibnamefont
  {Albataineh}}, \bibinfo {author} {\bibfnamefont {D.}~\bibnamefont {Androic}},
  \bibinfo {author} {\bibfnamefont {K.~A.}\ \bibnamefont {Aniol}}, \bibinfo
  {author} {\bibfnamefont {D.~S.}\ \bibnamefont {Armstrong}}, \bibinfo {author}
  {\bibfnamefont {T.}~\bibnamefont {Averett}}, \bibinfo {author} {\bibfnamefont
  {C.}~\bibnamefont {Ayerbe~Gayoso}}, \bibinfo {author} {\bibfnamefont {S.~K.}\
  \bibnamefont {Barcus}}, \bibinfo {author} {\bibfnamefont {V.}~\bibnamefont
  {Bellini}}, \bibinfo {author} {\bibfnamefont {R.~S.}\ \bibnamefont
  {Beminiwattha}}, \bibinfo {author} {\bibfnamefont {J.~F.}\ \bibnamefont
  {Benesch}}, \bibinfo {author} {\bibfnamefont {H.}~\bibnamefont {Bhatt}},
  \bibinfo {author} {\bibfnamefont {D.}~\bibnamefont {Bhatta~Pathak}}, \bibinfo
  {author} {\bibfnamefont {D.}~\bibnamefont {Bhetuwal}}, \bibinfo {author}
  {\bibfnamefont {B.}~\bibnamefont {Blaikie}}\ \emph {et~al.} (\bibinfo
  {collaboration} {CREX Collaboration}),\ }\href
  {https://doi.org/10.1103/PhysRevLett.129.042501} {\bibfield  {journal}
  {\bibinfo  {journal} {Phys. Rev. Lett.}\ }\textbf {\bibinfo {volume} {129}},\
  \bibinfo {pages} {042501} (\bibinfo {year} {2022})}\BibitemShut {NoStop}%
\bibitem [{\citenamefont {Tsang}\ \emph {et~al.}(2012)\citenamefont {Tsang},
  \citenamefont {Stone}, \citenamefont {Camera}, \citenamefont {Danielewicz},
  \citenamefont {Gandolfi}, \citenamefont {Hebeler}, \citenamefont {Horowitz},
  \citenamefont {Lee}, \citenamefont {Lynch}, \citenamefont {Kohley},
  \citenamefont {Lemmon}, \citenamefont {M\"oller}, \citenamefont {Murakami},
  \citenamefont {Riordan}, \citenamefont {Roca-Maza}, \citenamefont
  {Sammarruca}, \citenamefont {Steiner}, \citenamefont {Vida\~na},\ and\
  \citenamefont {Yennello}}]{tsang2012}%
  \BibitemOpen
  \bibfield  {author} {\bibinfo {author} {\bibfnamefont {M.~B.}\ \bibnamefont
  {Tsang}}, \bibinfo {author} {\bibfnamefont {J.~R.}\ \bibnamefont {Stone}},
  \bibinfo {author} {\bibfnamefont {F.}~\bibnamefont {Camera}}, \bibinfo
  {author} {\bibfnamefont {P.}~\bibnamefont {Danielewicz}}, \bibinfo {author}
  {\bibfnamefont {S.}~\bibnamefont {Gandolfi}}, \bibinfo {author}
  {\bibfnamefont {K.}~\bibnamefont {Hebeler}}, \bibinfo {author} {\bibfnamefont
  {C.~J.}\ \bibnamefont {Horowitz}}, \bibinfo {author} {\bibfnamefont
  {J.}~\bibnamefont {Lee}}, \bibinfo {author} {\bibfnamefont {W.~G.}\
  \bibnamefont {Lynch}}, \bibinfo {author} {\bibfnamefont {Z.}~\bibnamefont
  {Kohley}}, \bibinfo {author} {\bibfnamefont {R.}~\bibnamefont {Lemmon}},
  \bibinfo {author} {\bibfnamefont {P.}~\bibnamefont {M\"oller}}, \bibinfo
  {author} {\bibfnamefont {T.}~\bibnamefont {Murakami}}, \bibinfo {author}
  {\bibfnamefont {S.}~\bibnamefont {Riordan}}, \bibinfo {author} {\bibfnamefont
  {X.}~\bibnamefont {Roca-Maza}}, \bibinfo {author} {\bibfnamefont
  {F.}~\bibnamefont {Sammarruca}}, \bibinfo {author} {\bibfnamefont {A.~W.}\
  \bibnamefont {Steiner}}, \bibinfo {author} {\bibfnamefont {I.}~\bibnamefont
  {Vida\~na}},\ and\ \bibinfo {author} {\bibfnamefont {S.~J.}\ \bibnamefont
  {Yennello}},\ }\href {https://doi.org/10.1103/PhysRevC.86.015803} {\bibfield
  {journal} {\bibinfo  {journal} {Phys. Rev. C}\ }\textbf {\bibinfo {volume}
  {86}},\ \bibinfo {pages} {015803} (\bibinfo {year} {2012})}\BibitemShut
  {NoStop}%
\bibitem [{\citenamefont {{Hagen}}\ \emph {et~al.}(2016)\citenamefont
  {{Hagen}}, \citenamefont {{Ekstr{\"o}m}}, \citenamefont {{Forss{\'e}n}},
  \citenamefont {{Jansen}}, \citenamefont {{Nazarewicz}}, \citenamefont
  {{Papenbrock}}, \citenamefont {{Wendt}}, \citenamefont {{Bacca}},
  \citenamefont {{Barnea}}, \citenamefont {{Carlsson}}, \citenamefont
  {{Drischler}}, \citenamefont {{Hebeler}}, \citenamefont {{Hjorth-Jensen}},
  \citenamefont {{Miorelli}}, \citenamefont {{Orlandini}}, \citenamefont
  {{Schwenk}},\ and\ \citenamefont {{Simonis}}}]{hagen2015}%
  \BibitemOpen
  \bibfield  {author} {\bibinfo {author} {\bibfnamefont {G.}~\bibnamefont
  {{Hagen}}}, \bibinfo {author} {\bibfnamefont {A.}~\bibnamefont
  {{Ekstr{\"o}m}}}, \bibinfo {author} {\bibfnamefont {C.}~\bibnamefont
  {{Forss{\'e}n}}}, \bibinfo {author} {\bibfnamefont {G.~R.}\ \bibnamefont
  {{Jansen}}}, \bibinfo {author} {\bibfnamefont {W.}~\bibnamefont
  {{Nazarewicz}}}, \bibinfo {author} {\bibfnamefont {T.}~\bibnamefont
  {{Papenbrock}}}, \bibinfo {author} {\bibfnamefont {K.~A.}\ \bibnamefont
  {{Wendt}}}, \bibinfo {author} {\bibfnamefont {S.}~\bibnamefont {{Bacca}}},
  \bibinfo {author} {\bibfnamefont {N.}~\bibnamefont {{Barnea}}}, \bibinfo
  {author} {\bibfnamefont {B.}~\bibnamefont {{Carlsson}}}, \bibinfo {author}
  {\bibfnamefont {C.}~\bibnamefont {{Drischler}}}, \bibinfo {author}
  {\bibfnamefont {K.}~\bibnamefont {{Hebeler}}}, \bibinfo {author}
  {\bibfnamefont {M.}~\bibnamefont {{Hjorth-Jensen}}}, \bibinfo {author}
  {\bibfnamefont {M.}~\bibnamefont {{Miorelli}}}, \bibinfo {author}
  {\bibfnamefont {G.}~\bibnamefont {{Orlandini}}}, \bibinfo {author}
  {\bibfnamefont {A.}~\bibnamefont {{Schwenk}}},\ and\ \bibinfo {author}
  {\bibfnamefont {J.}~\bibnamefont {{Simonis}}},\ }\href
  {https://doi.org/10.1038/nphys3529} {\bibfield  {journal} {\bibinfo
  {journal} {Nat. Phys.}\ }\textbf {\bibinfo {volume} {12}},\ \bibinfo {pages}
  {186} (\bibinfo {year} {2016})}\BibitemShut {NoStop}%
\bibitem [{\citenamefont {Brown}(2017)}]{brown2017}%
  \BibitemOpen
  \bibfield  {author} {\bibinfo {author} {\bibfnamefont {B.~A.}\ \bibnamefont
  {Brown}},\ }\href {https://doi.org/10.1103/PhysRevLett.119.122502} {\bibfield
   {journal} {\bibinfo  {journal} {Phys. Rev. Lett.}\ }\textbf {\bibinfo
  {volume} {119}},\ \bibinfo {pages} {122502} (\bibinfo {year}
  {2017})}\BibitemShut {NoStop}%
\bibitem [{\citenamefont {Fattoyev}\ \emph {et~al.}(2018)\citenamefont
  {Fattoyev}, \citenamefont {Piekarewicz},\ and\ \citenamefont
  {Horowitz}}]{fattoyev2018}%
  \BibitemOpen
  \bibfield  {author} {\bibinfo {author} {\bibfnamefont {F.~J.}\ \bibnamefont
  {Fattoyev}}, \bibinfo {author} {\bibfnamefont {J.}~\bibnamefont
  {Piekarewicz}},\ and\ \bibinfo {author} {\bibfnamefont {C.~J.}\ \bibnamefont
  {Horowitz}},\ }\href {https://doi.org/10.1103/PhysRevLett.120.172702}
  {\bibfield  {journal} {\bibinfo  {journal} {Phys. Rev. Lett.}\ }\textbf
  {\bibinfo {volume} {120}},\ \bibinfo {pages} {172702} (\bibinfo {year}
  {2018})}\BibitemShut {NoStop}%
\bibitem [{\citenamefont {Bertulani}\ and\ \citenamefont
  {Valencia}(2019)}]{bertulani2019}%
  \BibitemOpen
  \bibfield  {author} {\bibinfo {author} {\bibfnamefont {C.~A.}\ \bibnamefont
  {Bertulani}}\ and\ \bibinfo {author} {\bibfnamefont {J.}~\bibnamefont
  {Valencia}},\ }\href {https://doi.org/10.1103/PhysRevC.100.015802} {\bibfield
   {journal} {\bibinfo  {journal} {Phys. Rev. C}\ }\textbf {\bibinfo {volume}
  {100}},\ \bibinfo {pages} {015802} (\bibinfo {year} {2019})}\BibitemShut
  {NoStop}%
\bibitem [{\citenamefont {Koszor{\'u}s}\ \emph {et~al.}(2021)}]{koszorus2021}%
  \BibitemOpen
  \bibfield  {author} {\bibinfo {author} {\bibfnamefont {{\'A}.}~\bibnamefont
  {Koszor{\'u}s}}\ \emph {et~al.}\ }\href
  {https://doi.org/10.1038/s41567-020-01136-5} {\bibfield
  {journal} {\bibinfo  {journal} {Nat. Phys.}\ }\textbf {\bibinfo {volume}
  {17}},\ \bibinfo {pages} {439} (\bibinfo {year} {2021})}\BibitemShut
  {NoStop}%
\bibitem [{\citenamefont {{Garcia Ruiz}}\ \emph {et~al.}(2016)\citenamefont
  {{Garcia Ruiz}}, \citenamefont {{Bissell}}, \citenamefont {{Blaum}},
  \citenamefont {{Ekstr\"om}}, \citenamefont {{Fr\"ommgen}}, \citenamefont
  {{Hagen}}, \citenamefont {{Hammen}}, \citenamefont {{Hebeler}}, \citenamefont
  {{Holt}}, \citenamefont {{Jansen}}, \citenamefont {{Kowalska}}, \citenamefont
  {{Kreim}}, \citenamefont {{Nazarewicz}}, \citenamefont {{Neugart}},
  \citenamefont {{Neyens}}, \citenamefont {{N\"ortersh\"auser}}, \citenamefont
  {{Papenbrock}}, \citenamefont {{Papuga}}, \citenamefont {{Schwenk}},
  \citenamefont {{Simonis}}, \citenamefont {{Wendt}},\ and\ \citenamefont
  {{Yordanov}}}]{garciaruiz2016}%
  \BibitemOpen
  \bibfield  {author} {\bibinfo {author} {\bibfnamefont {R.~F.}~\bibnamefont
  {{Garcia Ruiz}}}\ \emph {et~al.}\ }\href
  {https://doi.org/10.1038/nphys3645} {\bibfield  {journal} {\bibinfo
  {journal} {Nat. Phys.}\ }\textbf {\bibinfo {volume} {12}},\ \bibinfo {pages}
  {594} (\bibinfo {year} {2016})}\BibitemShut {NoStop}%
\bibitem [{\citenamefont {{Miller}}\ \emph {et~al.}(2019)\citenamefont
  {{Miller}}, \citenamefont {{Minamisono}}, \citenamefont {{Klose}},
  \citenamefont {{Garand}}, \citenamefont {{Kujawa}}, \citenamefont {{Lantis}},
  \citenamefont {{Liu}}, \citenamefont {{Maa{\ss}}}, \citenamefont {{Mantica}},
  \citenamefont {{Nazarewicz}}, \citenamefont {{N{\"o}rtersh{\"a}user}},
  \citenamefont {{Pineda}}, \citenamefont {{Reinhard}}, \citenamefont
  {{Rossi}}, \citenamefont {{Sommer}}, \citenamefont {{Sumithrarachchi}},
  \citenamefont {{Teigelh{\"o}fer}},\ and\ \citenamefont
  {{Watkins}}}]{miller2019}%
  \BibitemOpen
  \bibfield  {author} {\bibinfo {author} {\bibfnamefont {A.~J.}~\bibnamefont
  {{Miller}}}\ \emph {et~al.}\ }\href
  {https://doi.org/10.1038/s41567-019-0416-9} {\bibfield  {journal} {\bibinfo
  {journal} {Nat. Phys.}\ }\textbf {\bibinfo {volume} {15}},\ \bibinfo {pages}
  {432} (\bibinfo {year} {2019})}\BibitemShut {NoStop}%
\bibitem [{\citenamefont {Malbrunot-Ettenauer}\ \emph
  {et~al.}(2022)\citenamefont {Malbrunot-Ettenauer}, \citenamefont {Kaufmann},
  \citenamefont {Bacca}, \citenamefont {Barbieri}, \citenamefont {Billowes},
  \citenamefont {Bissell}, \citenamefont {Blaum}, \citenamefont {Cheal},
  \citenamefont {Duguet}, \citenamefont {Ruiz}, \citenamefont {Gins},
  \citenamefont {Gorges}, \citenamefont {Hagen}, \citenamefont {Heylen},
  \citenamefont {Holt}, \citenamefont {Jansen}, \citenamefont
  {Kanellakopoulos}, \citenamefont {Kortelainen}, \citenamefont {Miyagi},
  \citenamefont {Navr\'atil}, \citenamefont {Nazarewicz}, \citenamefont
  {Neugart}, \citenamefont {Neyens}, \citenamefont {N\"ortersh\"auser},
  \citenamefont {Novario}, \citenamefont {Papenbrock}, \citenamefont
  {Ratajczyk}, \citenamefont {Reinhard}, \citenamefont {Rodr\'{\i}guez},
  \citenamefont {S\'anchez}, \citenamefont {Sailer}, \citenamefont {Schwenk},
  \citenamefont {Simonis}, \citenamefont {Som\`a}, \citenamefont {Stroberg},
  \citenamefont {Wehner}, \citenamefont {Wraith}, \citenamefont {Xie},
  \citenamefont {Xu}, \citenamefont {Yang},\ and\ \citenamefont
  {Yordanov}}]{malbrunot2022}%
  \BibitemOpen
  \bibfield  {author} {\bibinfo {author} {\bibfnamefont {S.}~\bibnamefont
  {{Malbrunot-Ettenauer}}}\ \emph {et~al.}\ }\href
  {https://doi.org/10.1103/PhysRevLett.128.022502} {\bibfield  {journal}
  {\bibinfo  {journal} {Phys. Rev. Lett.}\ }\textbf {\bibinfo {volume} {128}},\
  \bibinfo {pages} {022502} (\bibinfo {year} {2022})}\BibitemShut {NoStop}%
\bibitem [{\citenamefont {Bissell}\ \emph {et~al.}(2016)\citenamefont
  {Bissell}, \citenamefont {Carette}, \citenamefont {Flanagan}, \citenamefont
  {Vingerhoets}, \citenamefont {Billowes}, \citenamefont {Blaum}, \citenamefont
  {Cheal}, \citenamefont {Fritzsche}, \citenamefont {Godefroid}, \citenamefont
  {Kowalska}, \citenamefont {Kr\"amer}, \citenamefont {Neugart}, \citenamefont
  {Neyens}, \citenamefont {N\"ortersh\"auser},\ and\ \citenamefont
  {Yordanov}}]{bissell2016}%
  \BibitemOpen
  \bibfield  {author} {\bibinfo {author} {\bibfnamefont {M.~L.}\ \bibnamefont
  {Bissell}}, \bibinfo {author} {\bibfnamefont {T.}~\bibnamefont {Carette}},
  \bibinfo {author} {\bibfnamefont {K.~T.}\ \bibnamefont {Flanagan}}, \bibinfo
  {author} {\bibfnamefont {P.}~\bibnamefont {Vingerhoets}}, \bibinfo {author}
  {\bibfnamefont {J.}~\bibnamefont {Billowes}}, \bibinfo {author}
  {\bibfnamefont {K.}~\bibnamefont {Blaum}}, \bibinfo {author} {\bibfnamefont
  {B.}~\bibnamefont {Cheal}}, \bibinfo {author} {\bibfnamefont
  {S.}~\bibnamefont {Fritzsche}}, \bibinfo {author} {\bibfnamefont
  {M.}~\bibnamefont {Godefroid}}, \bibinfo {author} {\bibfnamefont
  {M.}~\bibnamefont {Kowalska}}, \bibinfo {author} {\bibfnamefont
  {J.}~\bibnamefont {Kr\"amer}}, \bibinfo {author} {\bibfnamefont
  {R.}~\bibnamefont {Neugart}}, \bibinfo {author} {\bibfnamefont
  {G.}~\bibnamefont {Neyens}}, \bibinfo {author} {\bibfnamefont
  {W.}~\bibnamefont {N\"ortersh\"auser}},\ and\ \bibinfo {author}
  {\bibfnamefont {D.~T.}\ \bibnamefont {Yordanov}},\ }\href
  {https://doi.org/10.1103/PhysRevC.93.064318} {\bibfield  {journal} {\bibinfo
  {journal} {Phys. Rev. C}\ }\textbf {\bibinfo {volume} {93}},\ \bibinfo
  {pages} {064318} (\bibinfo {year} {2016})}\BibitemShut {NoStop}%
\bibitem [{\citenamefont {{de Groote}}\ \emph {et~al.}(2020)\citenamefont {{de
  Groote}}, \citenamefont {{Billowes}}, \citenamefont {{Binnersley}},
  \citenamefont {{Bissell}}, \citenamefont {{Cocolios}}, \citenamefont {{Day
  Goodacre}}, \citenamefont {{Farooq-Smith}}, \citenamefont {{Fedorov}},
  \citenamefont {{Flanagan}}, \citenamefont {{Franchoo}}, \citenamefont
  {{Garcia Ruiz}}, \citenamefont {{Gins}}, \citenamefont {{Holt}},
  \citenamefont {{Koszor{\'u}s}}, \citenamefont {{Lynch}}, \citenamefont
  {{Miyagi}}, \citenamefont {{Nazarewicz}}, \citenamefont {{Neyens}},
  \citenamefont {{Reinhard}}, \citenamefont {{Rothe}}, \citenamefont
  {{Stroke}}, \citenamefont {{Vernon}}, \citenamefont {{Wendt}}, \citenamefont
  {{Wilkins}}, \citenamefont {{Xu}},\ and\ \citenamefont
  {{Yang}}}]{groote2020}%
  \BibitemOpen
  \bibfield  {author} {\bibinfo {author} {\bibfnamefont {R.~P.}~\bibnamefont
  {{de Groote}}}\ \emph {et~al.}\ }\href {https://doi.org/10.1038/s41567-020-0868-y}
  {\bibfield  {journal} {\bibinfo  {journal} {Nat. Phys.}\ }\textbf {\bibinfo
  {volume} {16}},\ \bibinfo {pages} {620} (\bibinfo {year} {2020})}\BibitemShut
  {NoStop}%
\bibitem [{\citenamefont {Reponen}\ \emph {et~al.}(2021)\citenamefont
  {Reponen}, \citenamefont {de~Groote}, \citenamefont {Al~Ayoubi},
  \citenamefont {Beliuskina}, \citenamefont {Bissell}, \citenamefont
  {Campbell}, \citenamefont {Ca{\~{n}}ete}, \citenamefont {Cheal},
  \citenamefont {Chrysalidis}, \citenamefont {Delafosse}, \citenamefont
  {de~Roubin}, \citenamefont {Devlin}, \citenamefont {Eronen}, \citenamefont
  {Garcia~Ruiz}, \citenamefont {Geldhof}, \citenamefont {Gins}, \citenamefont
  {Hukkanen}, \citenamefont {Imgram}, \citenamefont {Kankainen}, \citenamefont
  {Kortelainen}, \citenamefont {Koszor{\'u}s}, \citenamefont
  {Kujanp{\"a}{\"a}}, \citenamefont {Mathieson}, \citenamefont {Nesterenko},
  \citenamefont {Pohjalainen}, \citenamefont {Vil{\'e}n}, \citenamefont
  {Zadvornaya},\ and\ \citenamefont {Moore}}]{reponen2021}%
  \BibitemOpen
  \bibfield  {author} {\bibinfo {author} {\bibfnamefont {M.}~\bibnamefont
  {{Reponen}}}\ \emph {et~al.}\ }\href
  {https://doi.org/10.1038/s41467-021-24888-x} {\bibfield  {journal} {\bibinfo
  {journal} {Nat. Commun.}\ }\textbf {\bibinfo {volume} {12}},\ \bibinfo {pages}
  {4596} (\bibinfo {year} {2021})}\BibitemShut {NoStop}%
\bibitem [{\citenamefont {{De Jager}}\ \emph {et~al.}(1974)\citenamefont {{De
  Jager}}, \citenamefont {{De Vries}},\ and\ \citenamefont {{De
  Vries}}}]{dejager1974}%
  \BibitemOpen
  \bibfield  {author} {\bibinfo {author} {\bibfnamefont {C.}~\bibnamefont {{De
  Jager}}}, \bibinfo {author} {\bibfnamefont {H.}~\bibnamefont {{De Vries}}},\
  and\ \bibinfo {author} {\bibfnamefont {C.}~\bibnamefont {{De Vries}}},\
  }\href {https://doi.org/https://doi.org/10.1016/S0092-640X(74)80002-1}
  {\bibfield  {journal} {\bibinfo  {journal} {At. Data Nucl. Data
  Tables}\ }\textbf {\bibinfo {volume} {14}},\ \bibinfo {pages} {479} (\bibinfo
  {year} {1974})},\ \bibinfo {note} {nuclear Charge and Moment
  Distributions}\BibitemShut {NoStop}%
\bibitem [{\citenamefont {Angeli}\ and\ \citenamefont
  {Marinova}(2013)}]{angeli2013}%
  \BibitemOpen
  \bibfield  {author} {\bibinfo {author} {\bibfnamefont {I.}~\bibnamefont
  {Angeli}}\ and\ \bibinfo {author} {\bibfnamefont {K.}~\bibnamefont
  {Marinova}},\ }\href {https://doi.org/10.1016/j.adt.2011.12.006} {\bibfield
  {journal} {\bibinfo  {journal} {At. Data Nucl. Data Tables}\
  }\textbf {\bibinfo {volume} {99}},\ \bibinfo {pages} {69 } (\bibinfo {year}
  {2013})}\BibitemShut {NoStop}%
\bibitem [{\citenamefont {Piekarewicz}\ and\ \citenamefont
  {Weppner}(2006)}]{piekarewicz2006}%
  \BibitemOpen
  \bibfield  {author} {\bibinfo {author} {\bibfnamefont {J.}~\bibnamefont
  {Piekarewicz}}\ and\ \bibinfo {author} {\bibfnamefont {S.}~\bibnamefont
  {Weppner}},\ }\href
  {https://doi.org/https://doi.org/10.1016/j.nuclphysa.2006.08.004} {\bibfield
  {journal} {\bibinfo  {journal} {Nucl. Phys. A}}\textbf {\bibinfo {volume}
  {778}},\ \bibinfo {pages} {10} (\bibinfo {year} {2006})}\BibitemShut
  {NoStop}%
\bibitem [{\citenamefont {{Reinhard}}\ \emph {et~al.}(2022)\citenamefont
  {{Reinhard}}, \citenamefont {{Roca-Maza}},\ and\ \citenamefont
  {{Nazarewicz}}}]{reinhard2022}%
  \BibitemOpen
  \bibfield  {author} {\bibinfo {author} {\bibfnamefont {P.-G.}\ \bibnamefont
  {{Reinhard}}}, \bibinfo {author} {\bibfnamefont {X.}~\bibnamefont
  {{Roca-Maza}}},\ and\ \bibinfo {author} {\bibfnamefont {W.}~\bibnamefont
  {{Nazarewicz}}},\ }\href {https://link.aps.org/doi/10.1103/PhysRevLett.129.232501}
  {\bibfield {journal} {\bibinfo  {journal} {Phys. Rev. Lett.}\ }\textbf {\bibinfo
  {volume} {129}},\ \bibinfo {pages} {232501} (\bibinfo {year}
  {2022})} \BibitemShut {NoStop}%
\bibitem [{\citenamefont {Coester}(1958)}]{coester1958}%
  \BibitemOpen
  \bibfield  {author} {\bibinfo {author} {\bibfnamefont {F.}~\bibnamefont
  {Coester}},\ }\href {https://doi.org/10.1016/0029-5582(58)90280-3} {\bibfield
   {journal} {\bibinfo  {journal} {Nucl. Phys.}\ }\textbf {\bibinfo {volume}
  {7}},\ \bibinfo {pages} {421 } (\bibinfo {year} {1958})}\BibitemShut
  {NoStop}%
\bibitem [{\citenamefont {Coester}\ and\ \citenamefont
  {K{\"u}mmel}(1960)}]{coester1960}%
  \BibitemOpen
  \bibfield  {author} {\bibinfo {author} {\bibfnamefont {F.}~\bibnamefont
  {Coester}}\ and\ \bibinfo {author} {\bibfnamefont {H.}~\bibnamefont
  {K{\"u}mmel}},\ }\href {https://doi.org/10.1016/0029-5582(60)90140-1}
  {\bibfield  {journal} {\bibinfo  {journal} {Nucl. Phys.}\ }\textbf {\bibinfo
  {volume} {17}},\ \bibinfo {pages} {477 } (\bibinfo {year}
  {1960})}\BibitemShut {NoStop}%
\bibitem [{\citenamefont {{\v {C}}{\'\i}{\v z}ek}(1966)}]{cizek1966}%
  \BibitemOpen
  \bibfield  {author} {\bibinfo {author} {\bibfnamefont {J.}~\bibnamefont {{\v
  {C}}{\'\i}{\v z}ek}},\ }\href {https://doi.org/10.1063/1.1727484} {\bibfield
  {journal} {\bibinfo  {journal} {J. Chem. Phys.}\ }\textbf {\bibinfo {volume}
  {45}},\ \bibinfo {pages} {4256} (\bibinfo {year} {1966})}\BibitemShut
  {NoStop}%
\bibitem [{\citenamefont {{\v {C}}{\'\i}{\v z}ek}(2007)}]{cizek1969}%
  \BibitemOpen
  \bibfield  {author} {\bibinfo {author} {\bibfnamefont {J.}~\bibnamefont {{\v
  {C}}{\'\i}{\v z}ek}},\ }\bibinfo {title} {On the use of the cluster expansion
  and the technique of diagrams in calculations of correlation effects in atoms
  and molecules},\ in\ \href {} {\emph
  {\bibinfo {booktitle} {\textit{Advances in Chemical Physics}}}}\ (\bibinfo
  {publisher} {John Wiley \& Sons, Inc., New York},\ \bibinfo {year} {2007})\ pp.\
  \bibinfo {pages} {35--89}\BibitemShut {NoStop}%
\bibitem [{\citenamefont {K{\"u}mmel}\ \emph {et~al.}(1978)\citenamefont
  {K{\"u}mmel}, \citenamefont {L{\"u}hrmann},\ and\ \citenamefont
  {Zabolitzky}}]{kuemmel1978}%
  \BibitemOpen
  \bibfield  {author} {\bibinfo {author} {\bibfnamefont {H.}~\bibnamefont
  {K{\"u}mmel}}, \bibinfo {author} {\bibfnamefont {K.~H.}\ \bibnamefont
  {L{\"u}hrmann}},\ and\ \bibinfo {author} {\bibfnamefont {J.~G.}\ \bibnamefont
  {Zabolitzky}},\ }\href {https://doi.org/10.1016/0370-1573(78)90081-9}
  {\bibfield  {journal} {\bibinfo  {journal} {Phys. Rep.}\ }\textbf {\bibinfo
  {volume} {36}},\ \bibinfo {pages} {1 } (\bibinfo {year} {1978})}\BibitemShut
  {NoStop}%
\bibitem [{\citenamefont {Bartlett}\ and\ \citenamefont
  {Musia\l{}}(2007)}]{bartlett2007}%
  \BibitemOpen
  \bibfield  {author} {\bibinfo {author} {\bibfnamefont {R.~J.}\ \bibnamefont
  {Bartlett}}\ and\ \bibinfo {author} {\bibfnamefont {M.}~\bibnamefont
  {Musia\l{}}},\ }\href {https://doi.org/10.1103/RevModPhys.79.291} {\bibfield
  {journal} {\bibinfo  {journal} {Rev. Mod. Phys.}\ }\textbf {\bibinfo {volume}
  {79}},\ \bibinfo {pages} {291} (\bibinfo {year} {2007})}\BibitemShut
  {NoStop}%
\bibitem [{\citenamefont {Hagen}\ \emph {et~al.}(2014)\citenamefont {Hagen},
  \citenamefont {Papenbrock}, \citenamefont {Hjorth-Jensen},\ and\
  \citenamefont {Dean}}]{hagen2013c}%
  \BibitemOpen
  \bibfield  {author} {\bibinfo {author} {\bibfnamefont {G.}~\bibnamefont
  {Hagen}}, \bibinfo {author} {\bibfnamefont {T.}~\bibnamefont {Papenbrock}},
  \bibinfo {author} {\bibfnamefont {M.}~\bibnamefont {Hjorth-Jensen}},\ and\
  \bibinfo {author} {\bibfnamefont {D.~J.}\ \bibnamefont {Dean}},\ }\href
  {https://doi.org/10.1088/0034-4885/77/9/096302} {\bibfield  {journal}
  {\bibinfo  {journal} {Rep. Prog. Phys.}\ }\textbf {\bibinfo {volume} {77}},\
  \bibinfo {pages} {096302} (\bibinfo {year} {2014})}\BibitemShut {NoStop}%
\bibitem [{\citenamefont {Schmidt}\ and\ \citenamefont
  {Fantoni}(1999)}]{schmidt1999}%
  \BibitemOpen
  \bibfield  {author} {\bibinfo {author} {\bibfnamefont {K.~E.}\ \bibnamefont
  {Schmidt}}\ and\ \bibinfo {author} {\bibfnamefont {S.}~\bibnamefont
  {Fantoni}},\ }\href {https://doi.org/10.1016/S0370-2693(98)01522-6}
  {\bibfield  {journal} {\bibinfo  {journal} {Phys. Lett. B}\ }\textbf
  {\bibinfo {volume} {446}},\ \bibinfo {pages} {99} (\bibinfo {year}
  {1999})}\BibitemShut {NoStop}%
\bibitem [{\citenamefont {Carlson}\ \emph {et~al.}(2015)\citenamefont
  {Carlson}, \citenamefont {Gandolfi}, \citenamefont {Pederiva}, \citenamefont
  {Pieper}, \citenamefont {Schiavilla}, \citenamefont {Schmidt},\ and\
  \citenamefont {Wiringa}}]{carlson2015}%
  \BibitemOpen
  \bibfield  {author} {\bibinfo {author} {\bibfnamefont {J.}~\bibnamefont
  {Carlson}}, \bibinfo {author} {\bibfnamefont {S.}~\bibnamefont {Gandolfi}},
  \bibinfo {author} {\bibfnamefont {F.}~\bibnamefont {Pederiva}}, \bibinfo
  {author} {\bibfnamefont {S.~C.}\ \bibnamefont {Pieper}}, \bibinfo {author}
  {\bibfnamefont {R.}~\bibnamefont {Schiavilla}}, \bibinfo {author}
  {\bibfnamefont {K.~E.}\ \bibnamefont {Schmidt}},\ and\ \bibinfo {author}
  {\bibfnamefont {R.~B.}\ \bibnamefont {Wiringa}},\ }\href
  {https://doi.org/10.1103/RevModPhys.87.1067} {\bibfield  {journal} {\bibinfo
  {journal} {Rev. Mod. Phys.}\ }\textbf {\bibinfo {volume} {87}},\ \bibinfo
  {pages} {1067} (\bibinfo {year} {2015})}\BibitemShut {NoStop}%
\bibitem [{\citenamefont {Lonardoni}\ \emph
  {et~al.}(2018{\natexlab{a}})\citenamefont {Lonardoni}, \citenamefont
  {Gandolfi}, \citenamefont {Lynn}, \citenamefont {Petrie}, \citenamefont
  {Carlson}, \citenamefont {Schmidt},\ and\ \citenamefont
  {Schwenk}}]{lonardoni2018a}%
  \BibitemOpen
  \bibfield  {author} {\bibinfo {author} {\bibfnamefont {D.}~\bibnamefont
  {Lonardoni}}, \bibinfo {author} {\bibfnamefont {S.}~\bibnamefont {Gandolfi}},
  \bibinfo {author} {\bibfnamefont {J.~E.}\ \bibnamefont {Lynn}}, \bibinfo
  {author} {\bibfnamefont {C.}~\bibnamefont {Petrie}}, \bibinfo {author}
  {\bibfnamefont {J.}~\bibnamefont {Carlson}}, \bibinfo {author} {\bibfnamefont
  {K.~E.}\ \bibnamefont {Schmidt}},\ and\ \bibinfo {author} {\bibfnamefont
  {A.}~\bibnamefont {Schwenk}},\ }\href
  {https://doi.org/10.1103/PhysRevC.97.044318} {\bibfield  {journal} {\bibinfo
  {journal} {Phys. Rev. C}\ }\textbf {\bibinfo {volume} {97}},\ \bibinfo
  {pages} {044318} (\bibinfo {year} {2018}{\natexlab{a}})}\BibitemShut
  {NoStop}%
\bibitem [{\citenamefont {Epelbaum}\ \emph {et~al.}(2009)\citenamefont
  {Epelbaum}, \citenamefont {Hammer},\ and\ \citenamefont
  {Mei\ss{}ner}}]{epelbaum2009}%
  \BibitemOpen
  \bibfield  {author} {\bibinfo {author} {\bibfnamefont {E.}~\bibnamefont
  {Epelbaum}}, \bibinfo {author} {\bibfnamefont {H.-W.}\ \bibnamefont
  {Hammer}},\ and\ \bibinfo {author} {\bibfnamefont {U.-G.}\ \bibnamefont
  {Mei\ss{}ner}},\ }\href {https://doi.org/10.1103/RevModPhys.81.1773}
  {\bibfield  {journal} {\bibinfo  {journal} {Rev. Mod. Phys.}\ }\textbf
  {\bibinfo {volume} {81}},\ \bibinfo {pages} {1773} (\bibinfo {year}
  {2009})}\BibitemShut {NoStop}%
\bibitem [{\citenamefont {Machleidt}\ and\ \citenamefont
  {Entem}(2011)}]{machleidt2011}%
  \BibitemOpen
  \bibfield  {author} {\bibinfo {author} {\bibfnamefont {R.}~\bibnamefont
  {Machleidt}}\ and\ \bibinfo {author} {\bibfnamefont {D.}~\bibnamefont
  {Entem}},\ }\href {https://doi.org/10.1016/j.physrep.2011.02.001} {\bibfield
  {journal} {\bibinfo  {journal} {Phys. Rep.}\ }\textbf {\bibinfo {volume}
  {503}},\ \bibinfo {pages} {1 } (\bibinfo {year} {2011})}\BibitemShut
  {NoStop}%
\bibitem [{\citenamefont {Hebeler}\ \emph {et~al.}(2011)\citenamefont
  {Hebeler}, \citenamefont {Bogner}, \citenamefont {Furnstahl}, \citenamefont
  {Nogga},\ and\ \citenamefont {Schwenk}}]{hebeler2011}%
  \BibitemOpen
  \bibfield  {author} {\bibinfo {author} {\bibfnamefont {K.}~\bibnamefont
  {Hebeler}}, \bibinfo {author} {\bibfnamefont {S.~K.}\ \bibnamefont {Bogner}},
  \bibinfo {author} {\bibfnamefont {R.~J.}\ \bibnamefont {Furnstahl}}, \bibinfo
  {author} {\bibfnamefont {A.}~\bibnamefont {Nogga}},\ and\ \bibinfo {author}
  {\bibfnamefont {A.}~\bibnamefont {Schwenk}},\ }\href
  {https://doi.org/10.1103/PhysRevC.83.031301} {\bibfield  {journal} {\bibinfo
  {journal} {Phys. Rev. C}\ }\textbf {\bibinfo {volume} {83}},\ \bibinfo
  {pages} {031301(R)} (\bibinfo {year} {2011})}\BibitemShut {NoStop}%
\bibitem [{\citenamefont {Ekstr\"om}\ \emph {et~al.}(2013)\citenamefont
  {Ekstr\"om}, \citenamefont {Baardsen}, \citenamefont {Forss\'en},
  \citenamefont {Hagen}, \citenamefont {Hjorth-Jensen}, \citenamefont {Jansen},
  \citenamefont {Machleidt}, \citenamefont {Nazarewicz}, \citenamefont
  {Papenbrock}, \citenamefont {Sarich},\ and\ \citenamefont
  {Wild}}]{ekstrom2013}%
  \BibitemOpen
  \bibfield  {author} {\bibinfo {author} {\bibfnamefont {A.}~\bibnamefont
  {Ekstr\"om}}, \bibinfo {author} {\bibfnamefont {G.}~\bibnamefont {Baardsen}},
  \bibinfo {author} {\bibfnamefont {C.}~\bibnamefont {Forss\'en}}, \bibinfo
  {author} {\bibfnamefont {G.}~\bibnamefont {Hagen}}, \bibinfo {author}
  {\bibfnamefont {M.}~\bibnamefont {Hjorth-Jensen}}, \bibinfo {author}
  {\bibfnamefont {G.~R.}\ \bibnamefont {Jansen}}, \bibinfo {author}
  {\bibfnamefont {R.}~\bibnamefont {Machleidt}}, \bibinfo {author}
  {\bibfnamefont {W.}~\bibnamefont {Nazarewicz}}, \bibinfo {author}
  {\bibfnamefont {T.}~\bibnamefont {Papenbrock}}, \bibinfo {author}
  {\bibfnamefont {J.}~\bibnamefont {Sarich}},\ and\ \bibinfo {author}
  {\bibfnamefont {S.~M.}\ \bibnamefont {Wild}},\ }\href
  {https://doi.org/10.1103/PhysRevLett.110.192502} {\bibfield  {journal}
  {\bibinfo  {journal} {Phys. Rev. Lett.}\ }\textbf {\bibinfo {volume} {110}},\
  \bibinfo {pages} {192502} (\bibinfo {year} {2013})}\BibitemShut {NoStop}%
\bibitem [{\citenamefont {Entem}\ \emph {et~al.}(2015)\citenamefont {Entem},
  \citenamefont {Kaiser}, \citenamefont {Machleidt},\ and\ \citenamefont
  {Nosyk}}]{entem2015}%
  \BibitemOpen
  \bibfield  {author} {\bibinfo {author} {\bibfnamefont {D.~R.}\ \bibnamefont
  {Entem}}, \bibinfo {author} {\bibfnamefont {N.}~\bibnamefont {Kaiser}},
  \bibinfo {author} {\bibfnamefont {R.}~\bibnamefont {Machleidt}},\ and\
  \bibinfo {author} {\bibfnamefont {Y.}~\bibnamefont {Nosyk}},\ }\href
  {https://doi.org/10.1103/PhysRevC.91.014002} {\bibfield  {journal} {\bibinfo
  {journal} {Phys. Rev. C}\ }\textbf {\bibinfo {volume} {91}},\ \bibinfo
  {pages} {014002} (\bibinfo {year} {2015})}\BibitemShut {NoStop}%
\bibitem [{\citenamefont {Epelbaum}\ \emph {et~al.}(2015)\citenamefont
  {Epelbaum}, \citenamefont {Krebs},\ and\ \citenamefont
  {Mei\ss{}ner}}]{epelbaum2015}%
  \BibitemOpen
  \bibfield  {author} {\bibinfo {author} {\bibfnamefont {E.}~\bibnamefont
  {Epelbaum}}, \bibinfo {author} {\bibfnamefont {H.}~\bibnamefont {Krebs}},\
  and\ \bibinfo {author} {\bibfnamefont {U.-G.}\ \bibnamefont {Mei\ss{}ner}},\
  }\href {https://doi.org/10.1103/PhysRevLett.115.122301} {\bibfield  {journal}
  {\bibinfo  {journal} {Phys. Rev. Lett.}\ }\textbf {\bibinfo {volume} {115}},\
  \bibinfo {pages} {122301} (\bibinfo {year} {2015})}\BibitemShut {NoStop}%
\bibitem [{\citenamefont {H{\"u}ther}\ \emph {et~al.}(2020)\citenamefont
  {H{\"u}ther}, \citenamefont {Vobig}, \citenamefont {Hebeler}, \citenamefont
  {Machleidt},\ and\ \citenamefont {Roth}}]{huther2020}%
  \BibitemOpen
  \bibfield  {author} {\bibinfo {author} {\bibfnamefont {T.}~\bibnamefont
  {H{\"u}ther}}, \bibinfo {author} {\bibfnamefont {K.}~\bibnamefont {Vobig}},
  \bibinfo {author} {\bibfnamefont {K.}~\bibnamefont {Hebeler}}, \bibinfo
  {author} {\bibfnamefont {R.}~\bibnamefont {Machleidt}},\ and\ \bibinfo
  {author} {\bibfnamefont {R.}~\bibnamefont {Roth}},\ }\href
  {https://doi.org/https://doi.org/10.1016/j.physletb.2020.135651} {\bibfield
  {journal} {\bibinfo  {journal} {Phys. Lett. B}\ }\textbf {\bibinfo {volume}
  {808}},\ \bibinfo {pages} {135651} (\bibinfo {year} {2020})}\BibitemShut
  {NoStop}%
\bibitem [{\citenamefont {Som\`a}\ \emph {et~al.}(2020)\citenamefont {Som\`a},
  \citenamefont {Navr\'atil}, \citenamefont {Raimondi}, \citenamefont
  {Barbieri},\ and\ \citenamefont {Duguet}}]{soma2020}%
  \BibitemOpen
  \bibfield  {author} {\bibinfo {author} {\bibfnamefont {V.}~\bibnamefont
  {Som\`a}}, \bibinfo {author} {\bibfnamefont {P.}~\bibnamefont {Navr\'atil}},
  \bibinfo {author} {\bibfnamefont {F.}~\bibnamefont {Raimondi}}, \bibinfo
  {author} {\bibfnamefont {C.}~\bibnamefont {Barbieri}},\ and\ \bibinfo
  {author} {\bibfnamefont {T.}~\bibnamefont {Duguet}},\ }\href
  {https://doi.org/10.1103/PhysRevC.101.014318} {\bibfield  {journal} {\bibinfo
   {journal} {Phys. Rev. C}\ }\textbf {\bibinfo {volume} {101}},\ \bibinfo
  {pages} {014318} (\bibinfo {year} {2020})}\BibitemShut {NoStop}%
\bibitem [{\citenamefont {Ekstr\"om}\ \emph {et~al.}(2015)\citenamefont
  {Ekstr\"om}, \citenamefont {Carlsson}, \citenamefont {Wendt}, \citenamefont
  {Forss{\'e}n}, \citenamefont {Jensen}, \citenamefont {Machleidt},\ and\
  \citenamefont {Wild}}]{ekstrom2015}%
  \BibitemOpen
  \bibfield  {author} {\bibinfo {author} {\bibfnamefont {A.}~\bibnamefont
  {Ekstr\"om}}, \bibinfo {author} {\bibfnamefont {B.~D.}\ \bibnamefont
  {Carlsson}}, \bibinfo {author} {\bibfnamefont {K.~A.}\ \bibnamefont {Wendt}},
  \bibinfo {author} {\bibfnamefont {C.}~\bibnamefont {Forss{\'e}n}}, \bibinfo
  {author} {\bibfnamefont {M.~H.}\ \bibnamefont {Jensen}}, \bibinfo {author}
  {\bibfnamefont {R.}~\bibnamefont {Machleidt}},\ and\ \bibinfo {author}
  {\bibfnamefont {S.~M.}\ \bibnamefont {Wild}},\ }\href
  {https://doi.org/10.1088/0954-3899/42/3/034003} {\bibfield  {journal}
  {\bibinfo  {journal} {J. Phys, G: Nucl. Part. Phys.}\ }\textbf {\bibinfo
  {volume} {42}},\ \bibinfo {pages} {034003} (\bibinfo {year}
  {2015})}\BibitemShut {NoStop}%
\bibitem [{\citenamefont {Payne}\ \emph {et~al.}(2019)\citenamefont {Payne},
  \citenamefont {Bacca}, \citenamefont {Hagen}, \citenamefont {Jiang},\ and\
  \citenamefont {Papenbrock}}]{payne2019}%
  \BibitemOpen
  \bibfield  {author} {\bibinfo {author} {\bibfnamefont {C.~G.}\ \bibnamefont
  {Payne}}, \bibinfo {author} {\bibfnamefont {S.}~\bibnamefont {Bacca}},
  \bibinfo {author} {\bibfnamefont {G.}~\bibnamefont {Hagen}}, \bibinfo
  {author} {\bibfnamefont {W.~G.}\ \bibnamefont {Jiang}},\ and\ \bibinfo
  {author} {\bibfnamefont {T.}~\bibnamefont {Papenbrock}},\ }\href
  {https://doi.org/10.1103/PhysRevC.100.061304} {\bibfield  {journal} {\bibinfo
   {journal} {Phys. Rev. C}\ }\textbf {\bibinfo {volume} {100}},\ \bibinfo
  {pages} {061304} (\bibinfo {year} {2019})}\BibitemShut {NoStop}%
\bibitem [{\citenamefont {Bagchi}\ \emph {et~al.}(2020)\citenamefont {Bagchi},
  \citenamefont {Kanungo}, \citenamefont {Tanaka}, \citenamefont {Geissel},
  \citenamefont {Doornenbal}, \citenamefont {Horiuchi}, \citenamefont {Hagen},
  \citenamefont {Suzuki}, \citenamefont {Tsunoda}, \citenamefont {Ahn},
  \citenamefont {Baba}, \citenamefont {Behr}, \citenamefont {Browne},
  \citenamefont {Chen}, \citenamefont {Cort\'es}, \citenamefont {Estrad\'e},
  \citenamefont {Fukuda}, \citenamefont {Holl}, \citenamefont {Itahashi},
  \citenamefont {Iwasa}, \citenamefont {Jansen}, \citenamefont {Jiang},
  \citenamefont {Kaur}, \citenamefont {Macchiavelli}, \citenamefont
  {Matsumoto}, \citenamefont {Momiyama}, \citenamefont {Murray}, \citenamefont
  {Nakamura}, \citenamefont {Novario}, \citenamefont {Ong}, \citenamefont
  {Otsuka}, \citenamefont {Papenbrock}, \citenamefont {Paschalis},
  \citenamefont {Prochazka}, \citenamefont {Scheidenberger}, \citenamefont
  {Schrock}, \citenamefont {Shimizu}, \citenamefont {Steppenbeck},
  \citenamefont {Sakurai}, \citenamefont {Suzuki}, \citenamefont {Suzuki},
  \citenamefont {Takechi}, \citenamefont {Takeda}, \citenamefont {Takeuchi},
  \citenamefont {Taniuchi}, \citenamefont {Wimmer},\ and\ \citenamefont
  {Yoshida}}]{bagchi2020}%
  \BibitemOpen
  \bibfield  {author} {\bibinfo {author} {\bibfnamefont {S.}~\bibnamefont
  {{Bagchi}}}\ \emph {et~al.}\ }\href
  {https://doi.org/10.1103/PhysRevLett.124.222504} {\bibfield  {journal}
  {\bibinfo  {journal} {Phys. Rev. Lett.}\ }\textbf {\bibinfo {volume} {124}},\
  \bibinfo {pages} {222504} (\bibinfo {year} {2020})}\BibitemShut {NoStop}%
\bibitem [{\citenamefont {Jiang}\ \emph {et~al.}(2020)\citenamefont {Jiang},
  \citenamefont {Ekstr\"om}, \citenamefont {Forss\'en}, \citenamefont {Hagen},
  \citenamefont {Jansen},\ and\ \citenamefont {Papenbrock}}]{jiang2020}%
  \BibitemOpen
  \bibfield  {author} {\bibinfo {author} {\bibfnamefont {W.~G.}\ \bibnamefont
  {Jiang}}, \bibinfo {author} {\bibfnamefont {A.}~\bibnamefont {Ekstr\"om}},
  \bibinfo {author} {\bibfnamefont {C.}~\bibnamefont {Forss\'en}}, \bibinfo
  {author} {\bibfnamefont {G.}~\bibnamefont {Hagen}}, \bibinfo {author}
  {\bibfnamefont {G.~R.}\ \bibnamefont {Jansen}},\ and\ \bibinfo {author}
  {\bibfnamefont {T.}~\bibnamefont {Papenbrock}},\ }\href
  {https://doi.org/10.1103/PhysRevC.102.054301} {\bibfield  {journal} {\bibinfo
   {journal} {Phys. Rev. C}\ }\textbf {\bibinfo {volume} {102}},\ \bibinfo
  {pages} {054301} (\bibinfo {year} {2020})}\BibitemShut {NoStop}%
\bibitem [{\citenamefont {Tichai}\ \emph {et~al.}(2019)\citenamefont {Tichai},
  \citenamefont {M\"uller}, \citenamefont {Vobig},\ and\ \citenamefont
  {Roth}}]{tichai2019}%
  \BibitemOpen
  \bibfield  {author} {\bibinfo {author} {\bibfnamefont {A.}~\bibnamefont
  {Tichai}}, \bibinfo {author} {\bibfnamefont {J.}~\bibnamefont {M\"uller}},
  \bibinfo {author} {\bibfnamefont {K.}~\bibnamefont {Vobig}},\ and\ \bibinfo
  {author} {\bibfnamefont {R.}~\bibnamefont {Roth}},\ }\href
  {https://doi.org/10.1103/PhysRevC.99.034321} {\bibfield  {journal} {\bibinfo
  {journal} {Phys. Rev. C}\ }\textbf {\bibinfo {volume} {99}},\ \bibinfo
  {pages} {034321} (\bibinfo {year} {2019})}\BibitemShut {NoStop}%
\bibitem [{\citenamefont {Novario}\ \emph {et~al.}(2020)\citenamefont
  {Novario}, \citenamefont {Hagen}, \citenamefont {Jansen},\ and\ \citenamefont
  {Papenbrock}}]{novario2020}%
  \BibitemOpen
  \bibfield  {author} {\bibinfo {author} {\bibfnamefont {S.~J.}\ \bibnamefont
  {Novario}}, \bibinfo {author} {\bibfnamefont {G.}~\bibnamefont {Hagen}},
  \bibinfo {author} {\bibfnamefont {G.~R.}\ \bibnamefont {Jansen}},\ and\
  \bibinfo {author} {\bibfnamefont {T.}~\bibnamefont {Papenbrock}},\ }\href
  {https://doi.org/10.1103/PhysRevC.102.051303} {\bibfield  {journal} {\bibinfo
   {journal} {Phys. Rev. C}\ }\textbf {\bibinfo {volume} {102}},\ \bibinfo
  {pages} {051303(R)} (\bibinfo {year} {2020})}\BibitemShut {NoStop}%
\bibitem [{\citenamefont {Hagen}\ \emph {et~al.}(2007)\citenamefont {Hagen},
  \citenamefont {Papenbrock}, \citenamefont {Dean}, \citenamefont {Schwenk},
  \citenamefont {Nogga}, \citenamefont {W\l{}och},\ and\ \citenamefont
  {Piecuch}}]{hagen2007a}%
  \BibitemOpen
  \bibfield  {author} {\bibinfo {author} {\bibfnamefont {G.}~\bibnamefont
  {Hagen}}, \bibinfo {author} {\bibfnamefont {T.}~\bibnamefont {Papenbrock}},
  \bibinfo {author} {\bibfnamefont {D.~J.}\ \bibnamefont {Dean}}, \bibinfo
  {author} {\bibfnamefont {A.}~\bibnamefont {Schwenk}}, \bibinfo {author}
  {\bibfnamefont {A.}~\bibnamefont {Nogga}}, \bibinfo {author} {\bibfnamefont
  {M.}~\bibnamefont {W\l{}och}},\ and\ \bibinfo {author} {\bibfnamefont
  {P.}~\bibnamefont {Piecuch}},\ }\href
  {https://doi.org/10.1103/PhysRevC.76.034302} {\bibfield  {journal} {\bibinfo
  {journal} {Phys. Rev. C}\ }\textbf {\bibinfo {volume} {76}},\ \bibinfo
  {pages} {034302} (\bibinfo {year} {2007})}\BibitemShut {NoStop}%
\bibitem [{\citenamefont {Roth}\ \emph {et~al.}(2012)\citenamefont {Roth},
  \citenamefont {Binder}, \citenamefont {Vobig}, \citenamefont {Calci},
  \citenamefont {Langhammer},\ and\ \citenamefont {Navr\'atil}}]{roth2012}%
  \BibitemOpen
  \bibfield  {author} {\bibinfo {author} {\bibfnamefont {R.}~\bibnamefont
  {Roth}}, \bibinfo {author} {\bibfnamefont {S.}~\bibnamefont {Binder}},
  \bibinfo {author} {\bibfnamefont {K.}~\bibnamefont {Vobig}}, \bibinfo
  {author} {\bibfnamefont {A.}~\bibnamefont {Calci}}, \bibinfo {author}
  {\bibfnamefont {J.}~\bibnamefont {Langhammer}},\ and\ \bibinfo {author}
  {\bibfnamefont {P.}~\bibnamefont {Navr\'atil}},\ }\href
  {https://doi.org/10.1103/PhysRevLett.109.052501} {\bibfield  {journal}
  {\bibinfo  {journal} {Phys. Rev. Lett.}\ }\textbf {\bibinfo {volume} {109}},\
  \bibinfo {pages} {052501} (\bibinfo {year} {2012})}\BibitemShut {NoStop}%
\bibitem [{\citenamefont {Gezerlis}\ \emph {et~al.}(2013)\citenamefont
  {Gezerlis}, \citenamefont {Tews}, \citenamefont {Epelbaum}, \citenamefont
  {Gandolfi}, \citenamefont {Hebeler}, \citenamefont {Nogga},\ and\
  \citenamefont {Schwenk}}]{gezerlis2013}%
  \BibitemOpen
  \bibfield  {author} {\bibinfo {author} {\bibfnamefont {A.}~\bibnamefont
  {Gezerlis}}, \bibinfo {author} {\bibfnamefont {I.}~\bibnamefont {Tews}},
  \bibinfo {author} {\bibfnamefont {E.}~\bibnamefont {Epelbaum}}, \bibinfo
  {author} {\bibfnamefont {S.}~\bibnamefont {Gandolfi}}, \bibinfo {author}
  {\bibfnamefont {K.}~\bibnamefont {Hebeler}}, \bibinfo {author} {\bibfnamefont
  {A.}~\bibnamefont {Nogga}},\ and\ \bibinfo {author} {\bibfnamefont
  {A.}~\bibnamefont {Schwenk}},\ }\href
  {https://doi.org/10.1103/PhysRevLett.111.032501} {\bibfield  {journal}
  {\bibinfo  {journal} {Phys. Rev. Lett.}\ }\textbf {\bibinfo {volume} {111}},\
  \bibinfo {pages} {032501} (\bibinfo {year} {2013})}\BibitemShut {NoStop}%
\bibitem [{\citenamefont {Gezerlis}\ \emph {et~al.}(2014)\citenamefont
  {Gezerlis}, \citenamefont {Tews}, \citenamefont {Epelbaum}, \citenamefont
  {Freunek}, \citenamefont {Gandolfi}, \citenamefont {Hebeler}, \citenamefont
  {Nogga},\ and\ \citenamefont {Schwenk}}]{gezerlis2014}%
  \BibitemOpen
  \bibfield  {author} {\bibinfo {author} {\bibfnamefont {A.}~\bibnamefont
  {Gezerlis}}, \bibinfo {author} {\bibfnamefont {I.}~\bibnamefont {Tews}},
  \bibinfo {author} {\bibfnamefont {E.}~\bibnamefont {Epelbaum}}, \bibinfo
  {author} {\bibfnamefont {M.}~\bibnamefont {Freunek}}, \bibinfo {author}
  {\bibfnamefont {S.}~\bibnamefont {Gandolfi}}, \bibinfo {author}
  {\bibfnamefont {K.}~\bibnamefont {Hebeler}}, \bibinfo {author} {\bibfnamefont
  {A.}~\bibnamefont {Nogga}},\ and\ \bibinfo {author} {\bibfnamefont
  {A.}~\bibnamefont {Schwenk}},\ }\href
  {https://doi.org/10.1103/PhysRevC.90.054323} {\bibfield  {journal} {\bibinfo
  {journal} {Phys. Rev. C}\ }\textbf {\bibinfo {volume} {90}},\ \bibinfo
  {pages} {054323} (\bibinfo {year} {2014})}\BibitemShut {NoStop}%
\bibitem [{\citenamefont {Lynn}\ \emph {et~al.}(2016)\citenamefont {Lynn},
  \citenamefont {Tews}, \citenamefont {Carlson}, \citenamefont {Gandolfi},
  \citenamefont {Gezerlis}, \citenamefont {Schmidt},\ and\ \citenamefont
  {Schwenk}}]{lynn2016}%
  \BibitemOpen
  \bibfield  {author} {\bibinfo {author} {\bibfnamefont {J.~E.}\ \bibnamefont
  {Lynn}}, \bibinfo {author} {\bibfnamefont {I.}~\bibnamefont {Tews}}, \bibinfo
  {author} {\bibfnamefont {J.}~\bibnamefont {Carlson}}, \bibinfo {author}
  {\bibfnamefont {S.}~\bibnamefont {Gandolfi}}, \bibinfo {author}
  {\bibfnamefont {A.}~\bibnamefont {Gezerlis}}, \bibinfo {author}
  {\bibfnamefont {K.~E.}\ \bibnamefont {Schmidt}},\ and\ \bibinfo {author}
  {\bibfnamefont {A.}~\bibnamefont {Schwenk}},\ }\href
  {https://doi.org/10.1103/PhysRevLett.116.062501} {\bibfield  {journal}
  {\bibinfo  {journal} {Phys. Rev. Lett.}\ }\textbf {\bibinfo {volume} {116}},\
  \bibinfo {pages} {062501} (\bibinfo {year} {2016})}\BibitemShut {NoStop}%
\bibitem [{\citenamefont {Lynn}\ \emph {et~al.}(2014)\citenamefont {Lynn},
  \citenamefont {Carlson}, \citenamefont {Epelbaum}, \citenamefont {Gandolfi},
  \citenamefont {Gezerlis},\ and\ \citenamefont {Schwenk}}]{lynn2014}%
  \BibitemOpen
  \bibfield  {author} {\bibinfo {author} {\bibfnamefont {J.~E.}\ \bibnamefont
  {Lynn}}, \bibinfo {author} {\bibfnamefont {J.}~\bibnamefont {Carlson}},
  \bibinfo {author} {\bibfnamefont {E.}~\bibnamefont {Epelbaum}}, \bibinfo
  {author} {\bibfnamefont {S.}~\bibnamefont {Gandolfi}}, \bibinfo {author}
  {\bibfnamefont {A.}~\bibnamefont {Gezerlis}},\ and\ \bibinfo {author}
  {\bibfnamefont {A.}~\bibnamefont {Schwenk}},\ }\href
  {https://doi.org/10.1103/PhysRevLett.113.192501} {\bibfield  {journal}
  {\bibinfo  {journal} {Phys. Rev. Lett.}\ }\textbf {\bibinfo {volume} {113}},\
  \bibinfo {pages} {192501} (\bibinfo {year} {2014})}\BibitemShut {NoStop}%
\bibitem [{\citenamefont {Lynn}\ \emph {et~al.}(2017)\citenamefont {Lynn},
  \citenamefont {Tews}, \citenamefont {Carlson}, \citenamefont {Gandolfi},
  \citenamefont {Gezerlis}, \citenamefont {Schmidt},\ and\ \citenamefont
  {Schwenk}}]{lynn2017}%
  \BibitemOpen
  \bibfield  {author} {\bibinfo {author} {\bibfnamefont {J.~E.}\ \bibnamefont
  {Lynn}}, \bibinfo {author} {\bibfnamefont {I.}~\bibnamefont {Tews}}, \bibinfo
  {author} {\bibfnamefont {J.}~\bibnamefont {Carlson}}, \bibinfo {author}
  {\bibfnamefont {S.}~\bibnamefont {Gandolfi}}, \bibinfo {author}
  {\bibfnamefont {A.}~\bibnamefont {Gezerlis}}, \bibinfo {author}
  {\bibfnamefont {K.~E.}\ \bibnamefont {Schmidt}},\ and\ \bibinfo {author}
  {\bibfnamefont {A.}~\bibnamefont {Schwenk}},\ }\href
  {https://doi.org/10.1103/PhysRevC.96.054007} {\bibfield  {journal} {\bibinfo
  {journal} {Phys. Rev. C}\ }\textbf {\bibinfo {volume} {96}},\ \bibinfo
  {pages} {054007} (\bibinfo {year} {2017})}\BibitemShut {NoStop}%
\bibitem [{\citenamefont {Lonardoni}\ \emph
  {et~al.}(2018{\natexlab{b}})\citenamefont {Lonardoni}, \citenamefont
  {Carlson}, \citenamefont {Gandolfi}, \citenamefont {Lynn}, \citenamefont
  {Schmidt}, \citenamefont {Schwenk},\ and\ \citenamefont
  {Wang}}]{lonardoni2018}%
  \BibitemOpen
  \bibfield  {author} {\bibinfo {author} {\bibfnamefont {D.}~\bibnamefont
  {Lonardoni}}, \bibinfo {author} {\bibfnamefont {J.}~\bibnamefont {Carlson}},
  \bibinfo {author} {\bibfnamefont {S.}~\bibnamefont {Gandolfi}}, \bibinfo
  {author} {\bibfnamefont {J.~E.}\ \bibnamefont {Lynn}}, \bibinfo {author}
  {\bibfnamefont {K.~E.}\ \bibnamefont {Schmidt}}, \bibinfo {author}
  {\bibfnamefont {A.}~\bibnamefont {Schwenk}},\ and\ \bibinfo {author}
  {\bibfnamefont {X.~B.}\ \bibnamefont {Wang}},\ }\href
  {https://doi.org/10.1103/PhysRevLett.120.122502} {\bibfield  {journal}
  {\bibinfo  {journal} {Phys. Rev. Lett.}\ }\textbf {\bibinfo {volume} {120}},\
  \bibinfo {pages} {122502} (\bibinfo {year} {2018}{\natexlab{b}})}\BibitemShut
  {NoStop}%
\bibitem [{\citenamefont {Lonardoni}\ \emph
  {et~al.}(2018{\natexlab{c}})\citenamefont {Lonardoni}, \citenamefont
  {Gandolfi}, \citenamefont {Wang},\ and\ \citenamefont
  {Carlson}}]{lonardoni2018b}%
  \BibitemOpen
  \bibfield  {author} {\bibinfo {author} {\bibfnamefont {D.}~\bibnamefont
  {Lonardoni}}, \bibinfo {author} {\bibfnamefont {S.}~\bibnamefont {Gandolfi}},
  \bibinfo {author} {\bibfnamefont {X.~B.}\ \bibnamefont {Wang}},\ and\
  \bibinfo {author} {\bibfnamefont {J.}~\bibnamefont {Carlson}},\ }\href
  {https://doi.org/10.1103/PhysRevC.98.014322} {\bibfield  {journal} {\bibinfo
  {journal} {Phys. Rev. C}\ }\textbf {\bibinfo {volume} {98}},\ \bibinfo
  {pages} {014322} (\bibinfo {year} {2018}{\natexlab{c}})}\BibitemShut
  {NoStop}%
\bibitem [{\citenamefont {Lynn}\ \emph {et~al.}(2020)\citenamefont {Lynn},
  \citenamefont {Lonardoni}, \citenamefont {Carlson}, \citenamefont {Chen},
  \citenamefont {Detmold}, \citenamefont {Gandolfi},\ and\ \citenamefont
  {Schwenk}}]{lynn2019a}%
  \BibitemOpen
  \bibfield  {author} {\bibinfo {author} {\bibfnamefont {J.~E.}\ \bibnamefont
  {Lynn}}, \bibinfo {author} {\bibfnamefont {D.}~\bibnamefont {Lonardoni}},
  \bibinfo {author} {\bibfnamefont {J.}~\bibnamefont {Carlson}}, \bibinfo
  {author} {\bibfnamefont {J.-W.}\ \bibnamefont {Chen}}, \bibinfo {author}
  {\bibfnamefont {W.}~\bibnamefont {Detmold}}, \bibinfo {author} {\bibfnamefont
  {S.}~\bibnamefont {Gandolfi}},\ and\ \bibinfo {author} {\bibfnamefont
  {A.}~\bibnamefont {Schwenk}},\ }\href
  {https://doi.org/10.1088/1361-6471/ab6af7} {\bibfield  {journal} {\bibinfo
  {journal} {J. Phys. G: Nucl. Part. Phys.}\ }\textbf {\bibinfo {volume}
  {47}},\ \bibinfo {pages} {045109} (\bibinfo {year} {2020})}\BibitemShut
  {NoStop}%
\bibitem [{\citenamefont {Lim}\ \emph {et~al.}(2019)\citenamefont {Lim},
  \citenamefont {Carlson}, \citenamefont {Loizides}, \citenamefont {Lonardoni},
  \citenamefont {Lynn}, \citenamefont {Nagle}, \citenamefont {Orjuela~Koop},\
  and\ \citenamefont {Ouellette}}]{lim2019}%
  \BibitemOpen
  \bibfield  {author} {\bibinfo {author} {\bibfnamefont {S.~H.}\ \bibnamefont
  {Lim}}, \bibinfo {author} {\bibfnamefont {J.}~\bibnamefont {Carlson}},
  \bibinfo {author} {\bibfnamefont {C.}~\bibnamefont {Loizides}}, \bibinfo
  {author} {\bibfnamefont {D.}~\bibnamefont {Lonardoni}}, \bibinfo {author}
  {\bibfnamefont {J.~E.}\ \bibnamefont {Lynn}}, \bibinfo {author}
  {\bibfnamefont {J.~L.}\ \bibnamefont {Nagle}}, \bibinfo {author}
  {\bibfnamefont {J.~D.}\ \bibnamefont {Orjuela~Koop}},\ and\ \bibinfo {author}
  {\bibfnamefont {J.}~\bibnamefont {Ouellette}},\ }\href
  {https://doi.org/10.1103/PhysRevC.99.044904} {\bibfield  {journal} {\bibinfo
  {journal} {Phys. Rev. C}\ }\textbf {\bibinfo {volume} {99}},\ \bibinfo
  {pages} {044904} (\bibinfo {year} {2019})}\BibitemShut {NoStop}%
\bibitem [{\citenamefont {Cruz-Torres}\ \emph {et~al.}(2019)\citenamefont
  {Cruz-Torres}, \citenamefont {Li}, \citenamefont {Hauenstein}, \citenamefont
  {Schmidt}, \citenamefont {Nguyen}, \citenamefont {Abrams}, \citenamefont
  {Albataineh}, \citenamefont {Alsalmi}, \citenamefont {Androic}, \citenamefont
  {Aniol}, \citenamefont {Armstrong}, \citenamefont {Arrington}, \citenamefont
  {Atac}, \citenamefont {Averett}, \citenamefont {{Ayerbe Gayoso}}\ \emph
  {et~al.}}]{cruz2019}%
  \BibitemOpen
  \bibfield  {author} {\bibinfo {author} {\bibfnamefont {R.}~\bibnamefont
  {Cruz-Torres}}, \bibinfo {author} {\bibfnamefont {S.}~\bibnamefont {Li}},
  \bibinfo {author} {\bibfnamefont {F.}~\bibnamefont {Hauenstein}}, \bibinfo
  {author} {\bibfnamefont {A.}~\bibnamefont {Schmidt}}, \bibinfo {author}
  {\bibfnamefont {D.}~\bibnamefont {Nguyen}}, \bibinfo {author} {\bibfnamefont
  {D.}~\bibnamefont {Abrams}}, \bibinfo {author} {\bibfnamefont
  {H.}~\bibnamefont {Albataineh}}, \bibinfo {author} {\bibfnamefont
  {S.}~\bibnamefont {Alsalmi}}, \bibinfo {author} {\bibfnamefont
  {D.}~\bibnamefont {Androic}}, \bibinfo {author} {\bibfnamefont
  {K.}~\bibnamefont {Aniol}}, \bibinfo {author} {\bibfnamefont
  {W.}~\bibnamefont {Armstrong}}, \bibinfo {author} {\bibfnamefont
  {J.}~\bibnamefont {Arrington}}, \bibinfo {author} {\bibfnamefont
  {H.}~\bibnamefont {Atac}}, \bibinfo {author} {\bibfnamefont {T.}~\bibnamefont
  {Averett}}, \bibinfo {author} {\bibfnamefont {C.}~\bibnamefont {{Ayerbe
  Gayoso}}}\ \emph {et~al.},\ }\href
  {https://doi.org/https://doi.org/10.1016/j.physletb.2019.134890} {\bibfield
  {journal} {\bibinfo  {journal} {Phys. Lett. B}\ }\textbf {\bibinfo {volume}
  {797}},\ \bibinfo {pages} {134890} (\bibinfo {year} {2019})}\BibitemShut
  {NoStop}%
\bibitem [{\citenamefont {Cruz-Torres}\ \emph {et~al.}(2021)\citenamefont
  {Cruz-Torres}, \citenamefont {Lonardoni}, \citenamefont {Weiss},
  \citenamefont {Piarulli}, \citenamefont {Barnea}, \citenamefont
  {Higinbotham}, \citenamefont {Piasetzky}, \citenamefont {Schmidt},
  \citenamefont {Weinstein}, \citenamefont {Wiringa},\ and\ \citenamefont
  {Hen}}]{cruz2021}%
  \BibitemOpen
  \bibfield  {author} {\bibinfo {author} {\bibfnamefont {R.}~\bibnamefont
  {Cruz-Torres}}, \bibinfo {author} {\bibfnamefont {D.}~\bibnamefont
  {Lonardoni}}, \bibinfo {author} {\bibfnamefont {R.}~\bibnamefont {Weiss}},
  \bibinfo {author} {\bibfnamefont {M.}~\bibnamefont {Piarulli}}, \bibinfo
  {author} {\bibfnamefont {N.}~\bibnamefont {Barnea}}, \bibinfo {author}
  {\bibfnamefont {D.~W.}\ \bibnamefont {Higinbotham}}, \bibinfo {author}
  {\bibfnamefont {E.}~\bibnamefont {Piasetzky}}, \bibinfo {author}
  {\bibfnamefont {A.}~\bibnamefont {Schmidt}}, \bibinfo {author} {\bibfnamefont
  {L.~B.}\ \bibnamefont {Weinstein}}, \bibinfo {author} {\bibfnamefont {R.~B.}\
  \bibnamefont {Wiringa}},\ and\ \bibinfo {author} {\bibfnamefont
  {O.}~\bibnamefont {Hen}},\ }\href
  {https://doi.org/10.1038/s41567-020-01053-7} {\bibfield  {journal} {\bibinfo
  {journal} {Nat. Phys.}\ }\textbf {\bibinfo {volume} {17}},\ \bibinfo {pages}
  {306} (\bibinfo {year} {2021})}\BibitemShut {NoStop}%
\bibitem [{\citenamefont {Zhao}\ and\ \citenamefont
  {Gandolfi}(2016)}]{gandolfi2016}%
  \BibitemOpen
  \bibfield  {author} {\bibinfo {author} {\bibfnamefont {P.~W.}\ \bibnamefont
  {Zhao}}\ and\ \bibinfo {author} {\bibfnamefont {S.}~\bibnamefont
  {Gandolfi}},\ }\href {https://doi.org/10.1103/PhysRevC.94.041302} {\bibfield
  {journal} {\bibinfo  {journal} {Phys. Rev. C}\ }\textbf {\bibinfo {volume}
  {94}},\ \bibinfo {pages} {041302(R)} (\bibinfo {year} {2016})}\BibitemShut
  {NoStop}%
\bibitem [{\citenamefont {Klos}\ \emph {et~al.}(2016)\citenamefont {Klos},
  \citenamefont {Lynn}, \citenamefont {Tews}, \citenamefont {Gandolfi},
  \citenamefont {Gezerlis}, \citenamefont {Hammer}, \citenamefont
  {Hoferichter},\ and\ \citenamefont {Schwenk}}]{klos2016}%
  \BibitemOpen
  \bibfield  {author} {\bibinfo {author} {\bibfnamefont {P.}~\bibnamefont
  {Klos}}, \bibinfo {author} {\bibfnamefont {J.~E.}\ \bibnamefont {Lynn}},
  \bibinfo {author} {\bibfnamefont {I.}~\bibnamefont {Tews}}, \bibinfo {author}
  {\bibfnamefont {S.}~\bibnamefont {Gandolfi}}, \bibinfo {author}
  {\bibfnamefont {A.}~\bibnamefont {Gezerlis}}, \bibinfo {author}
  {\bibfnamefont {H.-W.}\ \bibnamefont {Hammer}}, \bibinfo {author}
  {\bibfnamefont {M.}~\bibnamefont {Hoferichter}},\ and\ \bibinfo {author}
  {\bibfnamefont {A.}~\bibnamefont {Schwenk}},\ }\href
  {https://doi.org/10.1103/PhysRevC.94.054005} {\bibfield  {journal} {\bibinfo
  {journal} {Phys. Rev. C}\ }\textbf {\bibinfo {volume} {94}},\ \bibinfo
  {pages} {054005} (\bibinfo {year} {2016})}\BibitemShut {NoStop}%
\bibitem [{\citenamefont {Gandolfi}\ \emph {et~al.}(2017)\citenamefont
  {Gandolfi}, \citenamefont {Hammer}, \citenamefont {Klos}, \citenamefont
  {Lynn},\ and\ \citenamefont {Schwenk}}]{gandolfi2017}%
  \BibitemOpen
  \bibfield  {author} {\bibinfo {author} {\bibfnamefont {S.}~\bibnamefont
  {Gandolfi}}, \bibinfo {author} {\bibfnamefont {H.-W.}\ \bibnamefont
  {Hammer}}, \bibinfo {author} {\bibfnamefont {P.}~\bibnamefont {Klos}},
  \bibinfo {author} {\bibfnamefont {J.~E.}\ \bibnamefont {Lynn}},\ and\
  \bibinfo {author} {\bibfnamefont {A.}~\bibnamefont {Schwenk}},\ }\href
  {https://doi.org/10.1103/PhysRevLett.118.232501} {\bibfield  {journal}
  {\bibinfo  {journal} {Phys. Rev. Lett.}\ }\textbf {\bibinfo {volume} {118}},\
  \bibinfo {pages} {232501} (\bibinfo {year} {2017})}\BibitemShut {NoStop}%
\bibitem [{\citenamefont {Tews}\ \emph {et~al.}(2016)\citenamefont {Tews},
  \citenamefont {Gandolfi}, \citenamefont {Gezerlis},\ and\ \citenamefont
  {Schwenk}}]{tews2016}%
  \BibitemOpen
  \bibfield  {author} {\bibinfo {author} {\bibfnamefont {I.}~\bibnamefont
  {Tews}}, \bibinfo {author} {\bibfnamefont {S.}~\bibnamefont {Gandolfi}},
  \bibinfo {author} {\bibfnamefont {A.}~\bibnamefont {Gezerlis}},\ and\
  \bibinfo {author} {\bibfnamefont {A.}~\bibnamefont {Schwenk}},\ }\href
  {https://doi.org/10.1103/PhysRevC.93.024305} {\bibfield  {journal} {\bibinfo
  {journal} {Phys. Rev. C}\ }\textbf {\bibinfo {volume} {93}},\ \bibinfo
  {pages} {024305} (\bibinfo {year} {2016})}\BibitemShut {NoStop}%
\bibitem [{\citenamefont {Buraczynski}\ and\ \citenamefont
  {Gezerlis}(2016)}]{buraczynski2016}%
  \BibitemOpen
  \bibfield  {author} {\bibinfo {author} {\bibfnamefont {M.}~\bibnamefont
  {Buraczynski}}\ and\ \bibinfo {author} {\bibfnamefont {A.}~\bibnamefont
  {Gezerlis}},\ }\href {https://doi.org/10.1103/PhysRevLett.116.152501}
  {\bibfield  {journal} {\bibinfo  {journal} {Phys. Rev. Lett.}\ }\textbf
  {\bibinfo {volume} {116}},\ \bibinfo {pages} {152501} (\bibinfo {year}
  {2016})}\BibitemShut {NoStop}%
\bibitem [{\citenamefont {Buraczynski}\ and\ \citenamefont
  {Gezerlis}(2017)}]{buraczynski2017}%
  \BibitemOpen
  \bibfield  {author} {\bibinfo {author} {\bibfnamefont {M.}~\bibnamefont
  {Buraczynski}}\ and\ \bibinfo {author} {\bibfnamefont {A.}~\bibnamefont
  {Gezerlis}},\ }\href {https://doi.org/10.1103/PhysRevC.95.044309} {\bibfield
  {journal} {\bibinfo  {journal} {Phys. Rev. C}\ }\textbf {\bibinfo {volume}
  {95}},\ \bibinfo {pages} {044309} (\bibinfo {year} {2017})}\BibitemShut
  {NoStop}%
\bibitem [{\citenamefont {Riz}\ \emph {et~al.}(2020{\natexlab{a}})\citenamefont
  {Riz}, \citenamefont {Gandolfi},\ and\ \citenamefont {Pederiva}}]{riz2018}%
  \BibitemOpen
  \bibfield  {author} {\bibinfo {author} {\bibfnamefont {L.}~\bibnamefont
  {Riz}}, \bibinfo {author} {\bibfnamefont {S.}~\bibnamefont {Gandolfi}},\ and\
  \bibinfo {author} {\bibfnamefont {F.}~\bibnamefont {Pederiva}},\ }\href
  {https://doi.org/10.1088/1361-6471/ab6520} {\bibfield  {journal} {\bibinfo
  {journal} {J. Phys. G}\ }\textbf {\bibinfo {volume}
  {47}},\ \bibinfo {pages} {045106} (\bibinfo {year}
  {2020}{\natexlab{a}})}\BibitemShut {NoStop}%
\bibitem [{\citenamefont {Tews}\ \emph
  {et~al.}(2018{\natexlab{a}})\citenamefont {Tews}, \citenamefont {Carlson},
  \citenamefont {Gandolfi},\ and\ \citenamefont {Reddy}}]{tews2018}%
  \BibitemOpen
  \bibfield  {author} {\bibinfo {author} {\bibfnamefont {I.}~\bibnamefont
  {Tews}}, \bibinfo {author} {\bibfnamefont {J.}~\bibnamefont {Carlson}},
  \bibinfo {author} {\bibfnamefont {S.}~\bibnamefont {Gandolfi}},\ and\
  \bibinfo {author} {\bibfnamefont {S.}~\bibnamefont {Reddy}},\ }\href
  {https://doi.org/10.3847/1538-4357/aac267} {\bibfield  {journal} {\bibinfo
  {journal} {Astrophys. J.}\ }\textbf {\bibinfo {volume} {860}},\ \bibinfo {pages} {149}
  (\bibinfo {year} {2018}{\natexlab{a}})}\BibitemShut {NoStop}%
\bibitem [{\citenamefont {Tews}\ \emph
  {et~al.}(2018{\natexlab{b}})\citenamefont {Tews}, \citenamefont {Margueron},\
  and\ \citenamefont {Reddy}}]{tews2018a}%
  \BibitemOpen
  \bibfield  {author} {\bibinfo {author} {\bibfnamefont {I.}~\bibnamefont
  {Tews}}, \bibinfo {author} {\bibfnamefont {J.}~\bibnamefont {Margueron}},\
  and\ \bibinfo {author} {\bibfnamefont {S.}~\bibnamefont {Reddy}},\ }\href
  {https://doi.org/10.1103/PhysRevC.98.045804} {\bibfield  {journal} {\bibinfo
  {journal} {Phys. Rev. C}\ }\textbf {\bibinfo {volume} {98}},\ \bibinfo
  {pages} {045804} (\bibinfo {year} {2018}{\natexlab{b}})}\BibitemShut
  {NoStop}%
\bibitem [{\citenamefont {Gandolfi}\ \emph {et~al.}(2018)\citenamefont
  {Gandolfi}, \citenamefont {Carlson}, \citenamefont {Roggero}, \citenamefont
  {Lynn},\ and\ \citenamefont {Reddy}}]{gandolfi2018}%
  \BibitemOpen
  \bibfield  {author} {\bibinfo {author} {\bibfnamefont {S.}~\bibnamefont
  {Gandolfi}}, \bibinfo {author} {\bibfnamefont {J.}~\bibnamefont {Carlson}},
  \bibinfo {author} {\bibfnamefont {A.}~\bibnamefont {Roggero}}, \bibinfo
  {author} {\bibfnamefont {J.~E.}\ \bibnamefont {Lynn}},\ and\ \bibinfo
  {author} {\bibfnamefont {S.}~\bibnamefont {Reddy}},\ }\href
  {https://doi.org/https://doi.org/10.1016/j.physletb.2018.07.073} {\bibfield
  {journal} {\bibinfo  {journal} {Phys. Lett. B}\ }\textbf {\bibinfo {volume}
  {785}},\ \bibinfo {pages} {232} (\bibinfo {year} {2018})}\BibitemShut
  {NoStop}%
\bibitem [{\citenamefont {Tews}\ \emph {et~al.}(2019)\citenamefont {Tews},
  \citenamefont {Margueron},\ and\ \citenamefont {Reddy}}]{tews2019}%
  \BibitemOpen
  \bibfield  {author} {\bibinfo {author} {\bibfnamefont {I.}~\bibnamefont
  {Tews}}, \bibinfo {author} {\bibfnamefont {J.}~\bibnamefont {Margueron}},\
  and\ \bibinfo {author} {\bibfnamefont {S.}~\bibnamefont {Reddy}},\ }\href
  {https://doi.org/10.1140/epja/i2019-12774-6} {\bibfield  {journal}
  {\bibinfo {journal} {Eur. Phys. J. A}\ }\textbf{\bibinfo {volume} {55}},\
  \bibinfo {pages} {97} (\bibinfo {year} {2019})}\BibitemShut {NoStop}%
\bibitem [{\citenamefont {Lonardoni}\ \emph {et~al.}(2020)\citenamefont
  {Lonardoni}, \citenamefont {Tews}, \citenamefont {Gandolfi},\ and\
  \citenamefont {Carlson}}]{Lonardoni:2020}%
  \BibitemOpen
  \bibfield  {author} {\bibinfo {author} {\bibfnamefont {D.}~\bibnamefont
  {Lonardoni}}, \bibinfo {author} {\bibfnamefont {I.}~\bibnamefont {Tews}},
  \bibinfo {author} {\bibfnamefont {S.}~\bibnamefont {Gandolfi}},\ and\
  \bibinfo {author} {\bibfnamefont {J.}~\bibnamefont {Carlson}},\ }\href
  {https://doi.org/10.1103/PhysRevResearch.2.022033} {\bibfield  {journal}
  {\bibinfo  {journal} {Phys. Rev. Res.}\ }\textbf {\bibinfo {volume}
  {2}},\ \bibinfo {pages} {022033(R)} (\bibinfo {year} {2020})}\BibitemShut
  {NoStop}%
\bibitem [{\citenamefont {Riz}\ \emph {et~al.}(2020{\natexlab{b}})\citenamefont
  {Riz}, \citenamefont {Pederiva}, \citenamefont {Lonardoni},\ and\
  \citenamefont {Gandolfi}}]{Riz2020}%
  \BibitemOpen
  \bibfield  {author} {\bibinfo {author} {\bibfnamefont {L.}~\bibnamefont
  {Riz}}, \bibinfo {author} {\bibfnamefont {F.}~\bibnamefont {Pederiva}},
  \bibinfo {author} {\bibfnamefont {D.}~\bibnamefont {Lonardoni}},\ and\
  \bibinfo {author} {\bibfnamefont {S.}~\bibnamefont {Gandolfi}},\ }\href
  {https://doi.org/10.3390/particles3040046} {\bibfield  {journal} {\bibinfo
  {journal} {Particles}\ }\textbf {\bibinfo {volume} {3}},\ \bibinfo {pages}
  {706} (\bibinfo {year} {2020}{\natexlab{b}})}\BibitemShut {NoStop}%
\bibitem [{\citenamefont {Hergert}(2020)}]{hergert2020}%
  \BibitemOpen
  \bibfield  {author} {\bibinfo {author} {\bibfnamefont {H.}~\bibnamefont
  {Hergert}},\ }\href {https://doi.org/10.3389/fphy.2020.00379} {\bibfield
  {journal} {\bibinfo {journal} {Front. Phys.}\ }\textbf {\bibinfo {volume}
  {8}},\ \bibinfo {pages} {379} (\bibinfo {year} {2020})}\BibitemShut {NoStop}%
\bibitem [{\citenamefont {Gandolfi}\ \emph {et~al.}(2020)\citenamefont
  {Gandolfi}, \citenamefont {Lonardoni}, \citenamefont {Lovato},\ and\
  \citenamefont {Piarulli}}]{gandolfi2020}%
  \BibitemOpen
  \bibfield  {author} {\bibinfo {author} {\bibfnamefont {S.}~\bibnamefont
  {Gandolfi}}, \bibinfo {author} {\bibfnamefont {D.}~\bibnamefont {Lonardoni}},
  \bibinfo {author} {\bibfnamefont {A.}~\bibnamefont {Lovato}},\ and\ \bibinfo
  {author} {\bibfnamefont {M.}~\bibnamefont {Piarulli}},\ }\href
  {https://doi.org/10.3389/fphy.2020.00117} {\bibfield {journal}
  {\bibinfo {journal} {Front. Phys.}\ }\textbf {\bibinfo {volume} {8}},\
  \bibinfo {pages} {117} (\bibinfo {year} {2020})}\BibitemShut {NoStop}%
\bibitem [{\citenamefont {Myers}\ and\ \citenamefont
  {Swiatecki}(1969)}]{Myers:1969}%
  \BibitemOpen
  \bibfield  {author} {\bibinfo {author} {\bibfnamefont {W.~D.}\ \bibnamefont
  {Myers}}\ and\ \bibinfo {author} {\bibfnamefont {W.}~\bibnamefont
  {Swiatecki}},\ }\href
  {https://doi.org/https://doi.org/10.1016/0003-4916(69)90202-4} {\bibfield
  {journal} {\bibinfo  {journal} {Ann. Phys. (N.Y.)}\ }\textbf {\bibinfo
  {volume} {55}},\ \bibinfo {pages} {395} (\bibinfo {year} {1969})}\BibitemShut
  {NoStop}%
\bibitem [{\citenamefont {Myers}\ and\ \citenamefont
  {Swiatecki}(1974)}]{Myers:1974}%
  \BibitemOpen
  \bibfield  {author} {\bibinfo {author} {\bibfnamefont {W.~D.}~\bibnamefont
  {Myers}}\ and\ \bibinfo {author} {\bibfnamefont {W.}~\bibnamefont
  {Swiatecki}},\ }\href
  {https://doi.org/https://doi.org/10.1016/0003-4916(74)90299-1} {\bibfield
  {journal} {\bibinfo  {journal} {Ann. Phys. (N.Y.)}\ }\textbf {\bibinfo
  {volume} {84}},\ \bibinfo {pages} {186} (\bibinfo {year} {1974})}\BibitemShut
  {NoStop}%
\bibitem [{\citenamefont {Myers}(1977)}]{Myers:1977}%
  \BibitemOpen
  \bibfield  {author} {\bibinfo {author} {\bibfnamefont {W.~D.}\ \bibnamefont
  {Myers}},\ }\href@noop {} {\emph {\bibinfo {title} {Droplet Model of Atomic
  Nuclei}}}\ (\bibinfo  {publisher} {IFI/Plenum Data Company},\ \bibinfo
  {address} {New York, USA},\ \bibinfo {year} {1977})\BibitemShut {NoStop}%
\bibitem [{\citenamefont {Myers}\ and\ \citenamefont
  {Swiatecki}(1980)}]{Myers:1980}%
  \BibitemOpen
  \bibfield  {author} {\bibinfo {author} {\bibfnamefont {W.~D.}~\bibnamefont
  {Myers}}\ and\ \bibinfo {author} {\bibfnamefont {W.}~\bibnamefont
  {Swiatecki}},\ }\href
  {https://doi.org/https://doi.org/10.1016/0375-9474(80)90623-5} {\bibfield
  {journal} {\bibinfo  {journal} {Nucl. Phys. A}}\textbf {\bibinfo
  {volume} {336}},\ \bibinfo {pages} {267} (\bibinfo {year}
  {1980})}\BibitemShut {NoStop}%
\bibitem [{\citenamefont {Pethick}\ and\ \citenamefont
  {Ravenhall}(1996)}]{pethick1996}%
  \BibitemOpen
  \bibfield  {author} {\bibinfo {author} {\bibfnamefont {C.~J.}\ \bibnamefont
  {Pethick}}\ and\ \bibinfo {author} {\bibfnamefont {D.~G.}\ \bibnamefont
  {Ravenhall}},\ }\href
  {https://doi.org/https://doi.org/10.1016/0375-9474(96)00216-3} {\bibfield
  {journal} {\bibinfo  {journal} {Nucl. Phys. A}}\textbf {\bibinfo {volume}
  {606}},\ \bibinfo {pages} {173} (\bibinfo {year} {1996})}\BibitemShut
  {NoStop}%
\bibitem [{\citenamefont {Kaufmann}\ \emph {et~al.}(2020)\citenamefont
  {Kaufmann}, \citenamefont {Simonis}, \citenamefont {Bacca}, \citenamefont
  {Billowes}, \citenamefont {Bissell}, \citenamefont {Blaum}, \citenamefont
  {Cheal}, \citenamefont {Ruiz}, \citenamefont {Gins}, \citenamefont {Gorges},
  \citenamefont {Hagen}, \citenamefont {Heylen}, \citenamefont
  {Kanellakopoulos}, \citenamefont {Malbrunot-Ettenauer}, \citenamefont
  {Miorelli}, \citenamefont {Neugart}, \citenamefont {Neyens}, \citenamefont
  {N\"ortersh\"auser}, \citenamefont {S\'anchez}, \citenamefont {Sailer},
  \citenamefont {Schwenk}, \citenamefont {Ratajczyk}, \citenamefont
  {Rodr\'{\i}guez}, \citenamefont {Wehner}, \citenamefont {Wraith},
  \citenamefont {Xie}, \citenamefont {Xu}, \citenamefont {Yang},\ and\
  \citenamefont {Yordanov}}]{kaufmann2020}%
  \BibitemOpen
  \bibfield  {author} {\bibinfo {author} {\bibfnamefont {S.}~\bibnamefont
  {{Kaufmann}}}\ \emph {et~al.}\ }\href {https://doi.org/10.1103/PhysRevLett.124.132502}
  {\bibfield  {journal} {\bibinfo  {journal} {Phys. Rev. Lett.}\ }\textbf
  {\bibinfo {volume} {124}},\ \bibinfo {pages} {132502} (\bibinfo {year}
  {2020})}\BibitemShut {NoStop}%
\bibitem [{\citenamefont {Miorelli}\ \emph {et~al.}(2018)\citenamefont
  {Miorelli}, \citenamefont {Bacca}, \citenamefont {Hagen},\ and\ \citenamefont
  {Papenbrock}}]{miorelli2018}%
  \BibitemOpen
  \bibfield  {author} {\bibinfo {author} {\bibfnamefont {M.}~\bibnamefont
  {Miorelli}}, \bibinfo {author} {\bibfnamefont {S.}~\bibnamefont {Bacca}},
  \bibinfo {author} {\bibfnamefont {G.}~\bibnamefont {Hagen}},\ and\ \bibinfo
  {author} {\bibfnamefont {T.}~\bibnamefont {Papenbrock}},\ }\href
  {https://doi.org/10.1103/PhysRevC.98.014324} {\bibfield  {journal} {\bibinfo
  {journal} {Phys. Rev. C}\ }\textbf {\bibinfo {volume} {98}},\ \bibinfo
  {pages} {014324} (\bibinfo {year} {2018})}\BibitemShut {NoStop}%
\bibitem [{\citenamefont {Hagen}\ \emph {et~al.}(2022)\citenamefont {Hagen},
  \citenamefont {Novario}, \citenamefont {Sun}, \citenamefont {Papenbrock},
  \citenamefont {Jansen}, \citenamefont {Lietz}, \citenamefont {Duguet},\ and\
  \citenamefont {Tichai}}]{hagen2022}%
  \BibitemOpen
  \bibfield  {author} {\bibinfo {author} {\bibfnamefont {G.}~\bibnamefont
  {Hagen}}, \bibinfo {author} {\bibfnamefont {S.~J.}\ \bibnamefont {Novario}},
  \bibinfo {author} {\bibfnamefont {Z.~H.}\ \bibnamefont {Sun}}, \bibinfo
  {author} {\bibfnamefont {T.}~\bibnamefont {Papenbrock}}, \bibinfo {author}
  {\bibfnamefont {G.~R.}\ \bibnamefont {Jansen}}, \bibinfo {author}
  {\bibfnamefont {J.~G.}\ \bibnamefont {Lietz}}, \bibinfo {author}
  {\bibfnamefont {T.}~\bibnamefont {Duguet}},\ and\ \bibinfo {author}
  {\bibfnamefont {A.}~\bibnamefont {Tichai}},\ }\href
  {https://link.aps.org/doi/10.1103/PhysRevC.105.064311} {\bibfield {journal}
  {\bibinfo {journal} {Phys. Rev. C}\ }\textbf {\bibinfo {volume} {105}},\
  \bibinfo {pages} {064311} (\bibinfo {year} {2022})}\BibitemShut {NoStop}%
\bibitem [{sup()}]{supp}%
  \BibitemOpen
  \href@noop {} {\bibinfo {title} {See {Supplemental Material} below for
  additional results and derivations of {LDM} corrections, which includes
  {Refs.}~\cite{wang2017, angeli2013, bartlett2007, hagen2009b, Myers:1969,
  Myers:1974, Myers:1977, Myers:1980}}}\BibitemShut {NoStop}%
\bibitem [{\citenamefont {Trzci\ifmmode~\acute{n}\else \'{n}\fi{}ska}\ \emph
  {et~al.}(2001)\citenamefont {Trzci\ifmmode~\acute{n}\else \'{n}\fi{}ska},
  \citenamefont {Jastrz\ifmmode~\mbox{\c{e}}\else \c{e}\fi{}bski},
  \citenamefont {Lubi\ifmmode~\acute{n}\else \'{n}\fi{}ski}, \citenamefont
  {Hartmann}, \citenamefont {Schmidt}, \citenamefont {von Egidy},\ and\
  \citenamefont {K\l{}os}}]{trzcinska2001}%
  \BibitemOpen
  \bibfield  {author} {\bibinfo {author} {\bibfnamefont {A.}~\bibnamefont
  {Trzci\ifmmode~\acute{n}\else \'{n}\fi{}ska}}, \bibinfo {author}
  {\bibfnamefont {J.}~\bibnamefont {Jastrz\ifmmode~\mbox{\c{e}}\else
  \c{e}\fi{}bski}}, \bibinfo {author} {\bibfnamefont {P.}~\bibnamefont
  {Lubi\ifmmode~\acute{n}\else \'{n}\fi{}ski}}, \bibinfo {author}
  {\bibfnamefont {F.~J.}\ \bibnamefont {Hartmann}}, \bibinfo {author}
  {\bibfnamefont {R.}~\bibnamefont {Schmidt}}, \bibinfo {author} {\bibfnamefont
  {T.}~\bibnamefont {von Egidy}},\ and\ \bibinfo {author} {\bibfnamefont
  {B.}~\bibnamefont {K\l{}os}},\ }\href
  {https://doi.org/10.1103/PhysRevLett.87.082501} {\bibfield  {journal}
  {\bibinfo  {journal} {Phys. Rev. Lett.}\ }\textbf {\bibinfo {volume} {87}},\
  \bibinfo {pages} {082501} (\bibinfo {year} {2001})}\BibitemShut {NoStop}%
\bibitem [{\citenamefont {Jastrz\ifmmode~\mbox{\c{e}}\else \c{e}\fi{}bski}\
  \emph {et~al.}(2004)\citenamefont {Jastrz\ifmmode~\mbox{\c{e}}\else
  \c{e}\fi{}bski}, \citenamefont {Trzci\ifmmode~\acute{n}\else \'{n}\fi{}ska},
  \citenamefont {Lubi\ifmmode~\acute{n}\else \'{n}\fi{}ski}, \citenamefont
  {K\l{}os}, \citenamefont {Hartmann}, \citenamefont {von Egidy},\ and\
  \citenamefont {Wycech}}]{jastrzebbski2004}%
  \BibitemOpen
  \bibfield  {author} {\bibinfo {author} {\bibfnamefont {J.}~\bibnamefont
  {Jastrz\ifmmode~\mbox{\c{e}}\else \c{e}\fi{}bski}}, \bibinfo {author}
  {\bibfnamefont {A.}~\bibnamefont {Trzci\ifmmode~\acute{n}\else
  \'{n}\fi{}ska}}, \bibinfo {author} {\bibfnamefont {P.}~\bibnamefont
  {Lubi\ifmmode~\acute{n}\else \'{n}\fi{}ski}}, \bibinfo {author}
  {\bibfnamefont {B.}~\bibnamefont {K\l{}os}}, \bibinfo {author} {\bibfnamefont
  {F.~J.}\ \bibnamefont {Hartmann}}, \bibinfo {author} {\bibfnamefont
  {T.}~\bibnamefont {von Egidy}},\ and\ \bibinfo {author} {\bibfnamefont
  {S.}~\bibnamefont {Wycech}},\ }\href
  {https://doi.org/10.1142/S0218301304002168} {\bibfield  {journal} {\bibinfo
  {journal} {Int. J. Mod. Phys. E}\ }\textbf {\bibinfo {volume} {13}},\
  \bibinfo {pages} {343} (\bibinfo {year} {2004})}\BibitemShut {NoStop}%
\bibitem [{\citenamefont {Lapoux}\ \emph {et~al.}(2016)\citenamefont {Lapoux},
  \citenamefont {Som\`a}, \citenamefont {Barbieri}, \citenamefont {Hergert},
  \citenamefont {Holt},\ and\ \citenamefont {Stroberg}}]{lapoux2016}%
  \BibitemOpen
  \bibfield  {author} {\bibinfo {author} {\bibfnamefont {V.}~\bibnamefont
  {Lapoux}}, \bibinfo {author} {\bibfnamefont {V.}~\bibnamefont {Som\`a}},
  \bibinfo {author} {\bibfnamefont {C.}~\bibnamefont {Barbieri}}, \bibinfo
  {author} {\bibfnamefont {H.}~\bibnamefont {Hergert}}, \bibinfo {author}
  {\bibfnamefont {J.~D.}\ \bibnamefont {Holt}},\ and\ \bibinfo {author}
  {\bibfnamefont {S.~R.}\ \bibnamefont {Stroberg}},\ }\href
  {https://doi.org/10.1103/PhysRevLett.117.052501} {\bibfield  {journal}
  {\bibinfo  {journal} {Phys. Rev. Lett.}\ }\textbf {\bibinfo {volume} {117}},\
  \bibinfo {pages} {052501} (\bibinfo {year} {2016})}\BibitemShut {NoStop}%
\bibitem [{\citenamefont {Zenihiro}\ \emph {et~al.}(2018)\citenamefont
  {Zenihiro}, \citenamefont {Sakaguchi}, \citenamefont {Terashima},
  \citenamefont {Uesaka}, \citenamefont {Hagen}, \citenamefont {Itoh},
  \citenamefont {Murakami}, \citenamefont {Nakatsugawa}, \citenamefont
  {Ohnishi}, \citenamefont {Sagawa}, \citenamefont {Takeda}, \citenamefont
  {Uchida}, \citenamefont {Yoshida}, \citenamefont {Yoshida},\ and\
  \citenamefont {Yosoi}}]{zenihiro2018}%
  \BibitemOpen
  \bibfield  {author} {\bibinfo {author} {\bibfnamefont {J.}~\bibnamefont
  {Zenihiro}}, \bibinfo {author} {\bibfnamefont {H.}~\bibnamefont {Sakaguchi}},
  \bibinfo {author} {\bibfnamefont {S.}~\bibnamefont {Terashima}}, \bibinfo
  {author} {\bibfnamefont {T.}~\bibnamefont {Uesaka}}, \bibinfo {author}
  {\bibfnamefont {G.}~\bibnamefont {Hagen}}, \bibinfo {author} {\bibfnamefont
  {M.}~\bibnamefont {Itoh}}, \bibinfo {author} {\bibfnamefont {T.}~\bibnamefont
  {Murakami}}, \bibinfo {author} {\bibfnamefont {Y.}~\bibnamefont
  {Nakatsugawa}}, \bibinfo {author} {\bibfnamefont {T.}~\bibnamefont
  {Ohnishi}}, \bibinfo {author} {\bibfnamefont {H.}~\bibnamefont {Sagawa}},
  \bibinfo {author} {\bibfnamefont {H.}~\bibnamefont {Takeda}}, \bibinfo
  {author} {\bibfnamefont {M.}~\bibnamefont {Uchida}}, \bibinfo {author}
  {\bibfnamefont {H.~P.}\ \bibnamefont {Yoshida}}, \bibinfo {author}
  {\bibfnamefont {S.}~\bibnamefont {Yoshida}},\ and\ \bibinfo {author}
  {\bibfnamefont {M.}~\bibnamefont {Yosoi}},\ }\href
  {https://arxiv.org/abs/1810.11796} {\Eprint
  {https://arxiv.org/abs/1810.11796} {arXiv:1810.11796}}\BibitemShut {NoStop}%
\bibitem [{\citenamefont {Warda}\ \emph {et~al.}(2009)\citenamefont {Warda},
  \citenamefont {Vi\~nas}, \citenamefont {Roca-Maza},\ and\ \citenamefont
  {Centelles}}]{Warda:2009}%
  \BibitemOpen
  \bibfield  {author} {\bibinfo {author} {\bibfnamefont {M.}~\bibnamefont
  {Warda}}, \bibinfo {author} {\bibfnamefont {X.}~\bibnamefont {Vi\~nas}},
  \bibinfo {author} {\bibfnamefont {X.}~\bibnamefont {Roca-Maza}},\ and\
  \bibinfo {author} {\bibfnamefont {M.}~\bibnamefont {Centelles}},\ }\href
  {https://doi.org/10.1103/PhysRevC.80.024316} {\bibfield  {journal} {\bibinfo
  {journal} {Phys. Rev. C}\ }\textbf {\bibinfo {volume} {80}},\ \bibinfo
  {pages} {024316} (\bibinfo {year} {2009})}\BibitemShut {NoStop}%
\bibitem [{\citenamefont {Hu}\ \emph {et~al.}(2022)\citenamefont {Hu},
  \citenamefont {Jiang}, \citenamefont {Miyagi}, \citenamefont {Sun},
  \citenamefont {Ekstr{\"o}m}, \citenamefont {Forss{\'e}n}, \citenamefont
  {Hagen}, \citenamefont {Holt}, \citenamefont {Papenbrock}, \citenamefont
  {Stroberg},\ and\ \citenamefont {Vernon}}]{hu2022}%
  \BibitemOpen
  \bibfield  {author} {\bibinfo {author} {\bibfnamefont {B.}~\bibnamefont
  {Hu}}, \bibinfo {author} {\bibfnamefont {W.}~\bibnamefont {Jiang}}, \bibinfo
  {author} {\bibfnamefont {T.}~\bibnamefont {Miyagi}}, \bibinfo {author}
  {\bibfnamefont {Z.}~\bibnamefont {Sun}}, \bibinfo {author} {\bibfnamefont
  {A.}~\bibnamefont {Ekstr{\"o}m}}, \bibinfo {author} {\bibfnamefont
  {C.}~\bibnamefont {Forss{\'e}n}}, \bibinfo {author} {\bibfnamefont
  {G.}~\bibnamefont {Hagen}}, \bibinfo {author} {\bibfnamefont {J.~D.}\
  \bibnamefont {Holt}}, \bibinfo {author} {\bibfnamefont {T.}~\bibnamefont
  {Papenbrock}}, \bibinfo {author} {\bibfnamefont {S.~R.}\ \bibnamefont
  {Stroberg}},\ and\ \bibinfo {author} {\bibfnamefont {I.}~\bibnamefont
  {Vernon}},\ }\href {https://doi.org/10.1038/s41567-022-01715-8}
  {\bibfield {journal} {\bibinfo {journal} {Nat. Phys.}\ }\textbf {\bibinfo
  {volume} {18}},\ \bibinfo {pages} {1196} (\bibinfo {year} {2022})}
  \BibitemShut {NoStop}%
\bibitem [{\citenamefont {Pineda}\ \emph {et~al.}(2021)\citenamefont {Pineda},
  \citenamefont {K\"onig}, \citenamefont {Rossi}, \citenamefont {Brown},
  \citenamefont {Incorvati}, \citenamefont {Lantis}, \citenamefont
  {Minamisono}, \citenamefont {N\"ortersh\"auser}, \citenamefont {Piekarewicz},
  \citenamefont {Powel},\ and\ \citenamefont {Sommer}}]{pineda2021}%
  \BibitemOpen
  \bibfield  {author} {\bibinfo {author} {\bibfnamefont {S.~V.}\ \bibnamefont
  {Pineda}}, \bibinfo {author} {\bibfnamefont {K.}~\bibnamefont {K\"onig}},
  \bibinfo {author} {\bibfnamefont {D.~M.}\ \bibnamefont {Rossi}}, \bibinfo
  {author} {\bibfnamefont {B.~A.}\ \bibnamefont {Brown}}, \bibinfo {author}
  {\bibfnamefont {A.}~\bibnamefont {Incorvati}}, \bibinfo {author}
  {\bibfnamefont {J.}~\bibnamefont {Lantis}}, \bibinfo {author} {\bibfnamefont
  {K.}~\bibnamefont {Minamisono}}, \bibinfo {author} {\bibfnamefont
  {W.}~\bibnamefont {N\"ortersh\"auser}}, \bibinfo {author} {\bibfnamefont
  {J.}~\bibnamefont {Piekarewicz}}, \bibinfo {author} {\bibfnamefont
  {R.}~\bibnamefont {Powel}},\ and\ \bibinfo {author} {\bibfnamefont
  {F.}~\bibnamefont {Sommer}},\ }\href
  {https://doi.org/10.1103/PhysRevLett.127.182503} {\bibfield  {journal}
  {\bibinfo  {journal} {Phys. Rev. Lett.}\ }\textbf {\bibinfo {volume} {127}},\
  \bibinfo {pages} {182503} (\bibinfo {year} {2021})}\BibitemShut {NoStop}%
\bibitem [{\citenamefont {Holt}\ \emph {et~al.}(2013)\citenamefont {Holt},
  \citenamefont {Men\'endez},\ and\ \citenamefont {Schwenk}}]{holt2013b}%
  \BibitemOpen
  \bibfield  {author} {\bibinfo {author} {\bibfnamefont {J.~D.}\ \bibnamefont
  {Holt}}, \bibinfo {author} {\bibfnamefont {J.}~\bibnamefont {Men\'endez}},\
  and\ \bibinfo {author} {\bibfnamefont {A.}~\bibnamefont {Schwenk}},\ }\href
  {https://doi.org/10.1103/PhysRevLett.110.022502} {\bibfield  {journal}
  {\bibinfo  {journal} {Phys. Rev. Lett.}\ }\textbf {\bibinfo {volume} {110}},\
  \bibinfo {pages} {022502} (\bibinfo {year} {2013})}\BibitemShut {NoStop}%
\bibitem [{\citenamefont {Morris}\ \emph {et~al.}(2018)\citenamefont {Morris},
  \citenamefont {Simonis}, \citenamefont {Stroberg}, \citenamefont {Stumpf},
  \citenamefont {Hagen}, \citenamefont {Holt}, \citenamefont {Jansen},
  \citenamefont {Papenbrock}, \citenamefont {Roth},\ and\ \citenamefont
  {Schwenk}}]{morris2018}%
  \BibitemOpen
  \bibfield  {author} {\bibinfo {author} {\bibfnamefont {T.~D.}\ \bibnamefont
  {Morris}}, \bibinfo {author} {\bibfnamefont {J.}~\bibnamefont {Simonis}},
  \bibinfo {author} {\bibfnamefont {S.~R.}\ \bibnamefont {Stroberg}}, \bibinfo
  {author} {\bibfnamefont {C.}~\bibnamefont {Stumpf}}, \bibinfo {author}
  {\bibfnamefont {G.}~\bibnamefont {Hagen}}, \bibinfo {author} {\bibfnamefont
  {J.~D.}\ \bibnamefont {Holt}}, \bibinfo {author} {\bibfnamefont {G.~R.}\
  \bibnamefont {Jansen}}, \bibinfo {author} {\bibfnamefont {T.}~\bibnamefont
  {Papenbrock}}, \bibinfo {author} {\bibfnamefont {R.}~\bibnamefont {Roth}},\
  and\ \bibinfo {author} {\bibfnamefont {A.}~\bibnamefont {Schwenk}},\ }\href
  {https://doi.org/10.1103/PhysRevLett.120.152503} {\bibfield  {journal}
  {\bibinfo  {journal} {Phys. Rev. Lett.}\ }\textbf {\bibinfo {volume} {120}},\
  \bibinfo {pages} {152503} (\bibinfo {year} {2018})}\BibitemShut {NoStop}%
\bibitem [{\citenamefont {Michel}\ \emph {et~al.}(2019)\citenamefont {Michel},
  \citenamefont {Li}, \citenamefont {Xu},\ and\ \citenamefont
  {Zuo}}]{Michel:2019}%
  \BibitemOpen
  \bibfield  {author} {\bibinfo {author} {\bibfnamefont {N.}~\bibnamefont
  {Michel}}, \bibinfo {author} {\bibfnamefont {J.~G.}\ \bibnamefont {Li}},
  \bibinfo {author} {\bibfnamefont {F.~R.}\ \bibnamefont {Xu}},\ and\ \bibinfo
  {author} {\bibfnamefont {W.}~\bibnamefont {Zuo}},\ }\href
  {https://doi.org/10.1103/PhysRevC.100.064303} {\bibfield  {journal} {\bibinfo
   {journal} {Phys. Rev. C}\ }\textbf {\bibinfo {volume} {100}},\ \bibinfo
  {pages} {064303} (\bibinfo {year} {2019})}\BibitemShut {NoStop}%
\bibitem [{\citenamefont {Randhawa}\ \emph {et~al.}(2019)\citenamefont
  {Randhawa}, \citenamefont {Kanungo}, \citenamefont {Holl}, \citenamefont
  {Holt}, \citenamefont {Navr\'atil}, \citenamefont {Strogerg}, \citenamefont
  {Hagen}, \citenamefont {Jansen}, \citenamefont {Alcorta}, \citenamefont
  {Andreolu}, \citenamefont {Barnes}, \citenamefont {Burbadge}, \citenamefont
  {Burke}, \citenamefont {Chen}, \citenamefont {Chester}, \citenamefont
  {Christian}, \citenamefont {Cruz}, \citenamefont {Davids}, \citenamefont
  {Even}, \citenamefont {Hackman}, \citenamefont {Henderson}, \citenamefont
  {Ishimoto}, \citenamefont {Jassal}, \citenamefont {Kaur}, \citenamefont
  {Keefe}, \citenamefont {Kisliuk}, \citenamefont {Kr\"ucken}, \citenamefont
  {Liang}, \citenamefont {Lighthall}, \citenamefont {McGee}, \citenamefont
  {Measures}, \citenamefont {Moukaddam}, \citenamefont {Padilla-Rodal},
  \citenamefont {Shotter}, \citenamefont {Thompson}, \citenamefont {Turko},
  \citenamefont {Williams},\ and\ \citenamefont {Workman}}]{Randhawa:2019}%
  \BibitemOpen
  \bibfield  {author} {\bibinfo {author} {\bibfnamefont {J.~S.}\ \bibnamefont
  {{Randhawa}}}\ \emph {et~al.}\ }\href {https://doi.org/10.1103/PhysRevC.99.021301}
  {\bibfield {journal} {\bibinfo {journal} {Phys. Rev. C}\ }\textbf {\bibinfo
  {volume} {99}},\ \bibinfo {pages} {021301} (\bibinfo {year} {2019})}
  \BibitemShut {NoStop}%
\bibitem [{\citenamefont {Stroberg}\ \emph {et~al.}(2021)\citenamefont
  {Stroberg}, \citenamefont {Holt}, \citenamefont {Schwenk},\ and\
  \citenamefont {Simonis}}]{Stroberg:2021}%
  \BibitemOpen
  \bibfield  {author} {\bibinfo {author} {\bibfnamefont {S.~R.}\ \bibnamefont
  {Stroberg}}, \bibinfo {author} {\bibfnamefont {J.~D.}\ \bibnamefont {Holt}},
  \bibinfo {author} {\bibfnamefont {A.}~\bibnamefont {Schwenk}},\ and\ \bibinfo
  {author} {\bibfnamefont {J.}~\bibnamefont {Simonis}},\ }\href
  {https://doi.org/10.1103/PhysRevLett.126.022501} {\bibfield  {journal}
  {\bibinfo  {journal} {Phys. Rev. Lett.}\ }\textbf {\bibinfo {volume} {126}},\
  \bibinfo {pages} {022501} (\bibinfo {year} {2021})}\BibitemShut {NoStop}%
\bibitem [{\citenamefont {Wang}\ \emph {et~al.}(2017)\citenamefont {Wang},
  \citenamefont {Audi}, \citenamefont {Kondev}, \citenamefont {Huang},
  \citenamefont {Naimi},\ and\ \citenamefont {Xu}}]{wang2017}%
  \BibitemOpen
  \bibfield  {author} {\bibinfo {author} {\bibfnamefont {M.}~\bibnamefont
  {Wang}}, \bibinfo {author} {\bibfnamefont {G.}~\bibnamefont {Audi}}, \bibinfo
  {author} {\bibfnamefont {F.~G.}\ \bibnamefont {Kondev}}, \bibinfo {author}
  {\bibfnamefont {W.}~\bibnamefont {Huang}}, \bibinfo {author} {\bibfnamefont
  {S.}~\bibnamefont {Naimi}},\ and\ \bibinfo {author} {\bibfnamefont
  {X.}~\bibnamefont {Xu}},\ }\href
  {https://doi.org/10.1088/1674-1137/41/3/030003} {\bibfield  {journal}
  {\bibinfo  {journal} {Chinese Physics C}\ }\textbf {\bibinfo {volume} {41}},\
  \bibinfo {pages} {030003} (\bibinfo {year} {2017})}\BibitemShut {NoStop}%
\bibitem [{\citenamefont {Pohl}\ \emph {et~al.}(2010)\citenamefont {Pohl},
  \citenamefont {Antognini}, \citenamefont {Nez}, \citenamefont {Amaro},
  \citenamefont {Biraben}, \citenamefont {Cardoso}, \citenamefont {Covita},
  \citenamefont {Dax}, \citenamefont {Dhawan}, \citenamefont {Fernandes},
  \citenamefont {Giesen}, \citenamefont {Graf}, \citenamefont {H{\"a}nsch},
  \citenamefont {Indelicato}, \citenamefont {Julien}\ \emph
  {et~al.}}]{pohl2010}%
  \BibitemOpen
  \bibfield  {author} {\bibinfo {author} {\bibfnamefont {R.}~\bibnamefont
  {Pohl}}, \bibinfo {author} {\bibfnamefont {A.}~\bibnamefont {Antognini}},
  \bibinfo {author} {\bibfnamefont {F.}~\bibnamefont {Nez}}, \bibinfo {author}
  {\bibfnamefont {F.~D.}\ \bibnamefont {Amaro}}, \bibinfo {author}
  {\bibfnamefont {F.}~\bibnamefont {Biraben}}, \bibinfo {author} {\bibfnamefont
  {J.~M.}\ \bibnamefont {Cardoso}}, \bibinfo {author} {\bibfnamefont {D.~S.}\
  \bibnamefont {Covita}}, \bibinfo {author} {\bibfnamefont {A.}~\bibnamefont
  {Dax}}, \bibinfo {author} {\bibfnamefont {S.}~\bibnamefont {Dhawan}},
  \bibinfo {author} {\bibfnamefont {L.~M.}\ \bibnamefont {Fernandes}}, \bibinfo
  {author} {\bibfnamefont {A.}~\bibnamefont {Giesen}}, \bibinfo {author}
  {\bibfnamefont {T.}~\bibnamefont {Graf}}, \bibinfo {author} {\bibfnamefont
  {T.~W.}\ \bibnamefont {H{\"a}nsch}}, \bibinfo {author} {\bibfnamefont
  {P.}~\bibnamefont {Indelicato}}, \bibinfo {author} {\bibfnamefont
  {L.}~\bibnamefont {Julien}}\ \emph {et~al.},\ }\href
  {https://doi.org/10.1038/nature09250} {\bibfield  {journal} {\bibinfo
  {journal} {Nature (London)}\ }\textbf {\bibinfo {volume} {466}},\ \bibinfo {pages}
  {213} (\bibinfo {year} {2010})}\BibitemShut {NoStop}%
\bibitem [{\citenamefont {Xiong}\ \emph {et~al.}(2019)\citenamefont {Xiong},
  \citenamefont {Gasparian}, \citenamefont {Gao}, \citenamefont {Dutta},
  \citenamefont {Khandaker}, \citenamefont {Liyanage}, \citenamefont {Pasyuk},
  \citenamefont {Peng}, \citenamefont {Bai}, \citenamefont {Ye}, \citenamefont
  {Gnanvo}, \citenamefont {Gu}, \citenamefont {Levillain}, \citenamefont {Yan},
  \citenamefont {Higinbotham}\ \emph {et~al.}}]{xiong2019}%
  \BibitemOpen
  \bibfield  {author} {\bibinfo {author} {\bibfnamefont {W.}~\bibnamefont
  {Xiong}}, \bibinfo {author} {\bibfnamefont {A.}~\bibnamefont {Gasparian}},
  \bibinfo {author} {\bibfnamefont {H.}~\bibnamefont {Gao}}, \bibinfo {author}
  {\bibfnamefont {D.}~\bibnamefont {Dutta}}, \bibinfo {author} {\bibfnamefont
  {M.}~\bibnamefont {Khandaker}}, \bibinfo {author} {\bibfnamefont
  {N.}~\bibnamefont {Liyanage}}, \bibinfo {author} {\bibfnamefont
  {E.}~\bibnamefont {Pasyuk}}, \bibinfo {author} {\bibfnamefont
  {C.}~\bibnamefont {Peng}}, \bibinfo {author} {\bibfnamefont {X.}~\bibnamefont
  {Bai}}, \bibinfo {author} {\bibfnamefont {L.}~\bibnamefont {Ye}}, \bibinfo
  {author} {\bibfnamefont {K.}~\bibnamefont {Gnanvo}}, \bibinfo {author}
  {\bibfnamefont {C.}~\bibnamefont {Gu}}, \bibinfo {author} {\bibfnamefont
  {M.}~\bibnamefont {Levillain}}, \bibinfo {author} {\bibfnamefont
  {X.}~\bibnamefont {Yan}}, \bibinfo {author} {\bibfnamefont {D.~W.}\
  \bibnamefont {Higinbotham}}\ \emph {et~al.},\ }\href
  {https://doi.org/10.1038/s41586-019-1721-2} {\bibfield  {journal} {\bibinfo
  {journal} {Nature (London)}\ }\textbf {\bibinfo {volume} {575}},\ \bibinfo {pages}
  {147} (\bibinfo {year} {2019})}\BibitemShut {NoStop}%
\bibitem [{\citenamefont {Filin}\ \emph {et~al.}(2020)\citenamefont {Filin},
  \citenamefont {Baru}, \citenamefont {Epelbaum}, \citenamefont {Krebs},
  \citenamefont {M\"oller},\ and\ \citenamefont {Reinert}}]{filin2020}%
  \BibitemOpen
  \bibfield  {author} {\bibinfo {author} {\bibfnamefont {A.~A.}\ \bibnamefont
  {Filin}}, \bibinfo {author} {\bibfnamefont {V.}~\bibnamefont {Baru}},
  \bibinfo {author} {\bibfnamefont {E.}~\bibnamefont {Epelbaum}}, \bibinfo
  {author} {\bibfnamefont {H.}~\bibnamefont {Krebs}}, \bibinfo {author}
  {\bibfnamefont {D.}~\bibnamefont {M\"oller}},\ and\ \bibinfo {author}
  {\bibfnamefont {P.}~\bibnamefont {Reinert}},\ }\href
  {https://doi.org/10.1103/PhysRevLett.124.082501} {\bibfield  {journal}
  {\bibinfo  {journal} {Phys. Rev. Lett.}\ }\textbf {\bibinfo {volume} {124}},\
  \bibinfo {pages} {082501} (\bibinfo {year} {2020})}\BibitemShut {NoStop}%
\bibitem [{\citenamefont {Naito}\ \emph {et~al.}(2022)\citenamefont {Naito},
  \citenamefont {Col\`o}, \citenamefont {Liang}, \citenamefont {Roca-Maza},\
  and\ \citenamefont {Sagawa}}]{naito2022}%
  \BibitemOpen
  \bibfield  {author} {\bibinfo {author} {\bibfnamefont {T.}~\bibnamefont
  {Naito}}, \bibinfo {author} {\bibfnamefont {G.}~\bibnamefont {Col\`o}},
  \bibinfo {author} {\bibfnamefont {H.}~\bibnamefont {Liang}}, \bibinfo
  {author} {\bibfnamefont {X.}~\bibnamefont {Roca-Maza}},\ and\ \bibinfo
  {author} {\bibfnamefont {H.}~\bibnamefont {Sagawa}},\ }\href
  {https://doi.org/10.1103/PhysRevC.105.L021304} {\bibfield  {journal}
  {\bibinfo  {journal} {Phys. Rev. C}\ }\textbf {\bibinfo {volume} {105}},\
  \bibinfo {pages} {L021304} (\bibinfo {year} {2022})}\BibitemShut {NoStop}%
\bibitem [{\citenamefont {Hagen}\ \emph {et~al.}(2009)\citenamefont {Hagen},
  \citenamefont {Papenbrock}, \citenamefont {Dean}, \citenamefont
  {Hjorth-Jensen},\ and\ \citenamefont {Asokan}}]{hagen2009b}%
  \BibitemOpen
  \bibfield  {author} {\bibinfo {author} {\bibfnamefont {G.}~\bibnamefont
  {Hagen}}, \bibinfo {author} {\bibfnamefont {T.}~\bibnamefont {Papenbrock}},
  \bibinfo {author} {\bibfnamefont {D.~J.}\ \bibnamefont {Dean}}, \bibinfo
  {author} {\bibfnamefont {M.}~\bibnamefont {Hjorth-Jensen}},\ and\ \bibinfo
  {author} {\bibfnamefont {B.~V.}\ \bibnamefont {Asokan}},\ }\href
  {https://doi.org/10.1103/PhysRevC.80.021306} {\bibfield  {journal} {\bibinfo
  {journal} {Phys. Rev. C}\ }\textbf {\bibinfo {volume} {80}},\ \bibinfo
  {pages} {021306(R)} (\bibinfo {year} {2009})}\BibitemShut {NoStop}%
\end{thebibliography}

%

\clearpage
\newpage

\appendix

\widowpenalty10000
\clubpenalty10000
\looseness=-2

\section{Supplemental Material: Trends of neutron skins and radii of mirror nuclei from first principles}

\subsection{Absolute binding energies and charge radii}

\begin{figure}[bth]
  \setlength\abovecaptionskip{-0.2\baselineskip}
  \hspace*{-8pt}
  \includegraphics[width=0.5\textwidth]{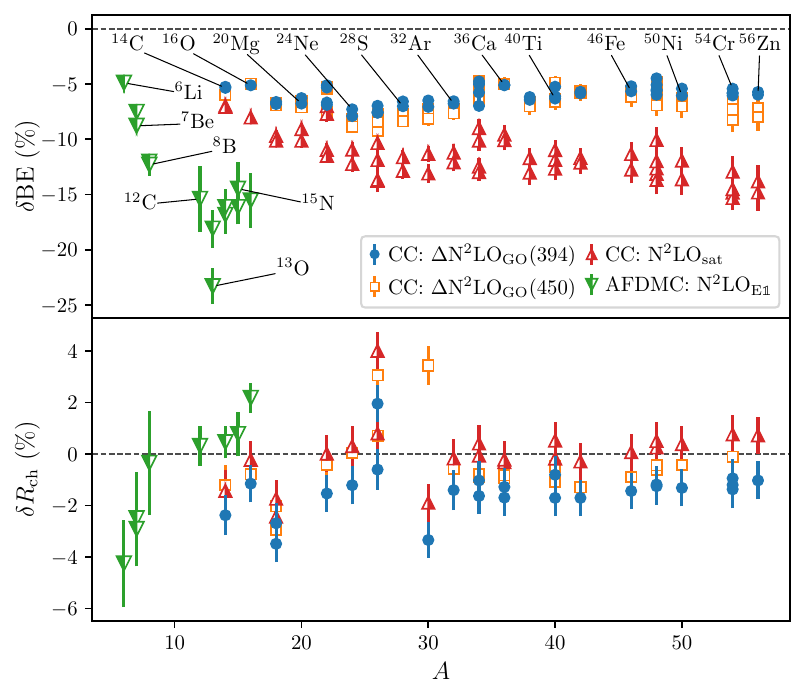}
  \caption{Top panel: Binding energies per nucleon residuals from
    \textit{ab initio} calculations compared to experimental data from
    Ref.~\cite{wang2017} plotted against mass number,
    $A$. Coupled-cluster results of nuclei with $14\le A\le 56$ using
    the $\Delta$N$^2$LO$_{\rm GO}$(394), $\Delta$N$^2$LO$_{\rm
      GO}$(450), and N$^2$LO$_{\rm sat}$(450) interactions are shown
    as solid blue circles, empty orange squares, and right-filled red
    triangles, respectively. Left-filled green triangles represent
    auxiliary field diffusion Monte Carlo results for $6\le A\le 18$
    using the N$^2$LO$_{E\mathbbm1}$ local chiral interaction. Select
    nuclei are labeled. Bottom panel: Charge radii residuals from
    \textit{ab initio} calculations compared to experimental results
    from Ref.~\cite{angeli2013} plotted against $A$.}
  \label{suppfig1}
\end{figure}

The mirror-difference quantities in Fig.~4, binding energy per nucleon
and charge radius, were computed from calculations of individual
mirror nuclei. However, because of their similar structures, these
mirror-differences tend to cancel deficiencies in the individual
quantities. These individual results, binding energy and charge
radius, are plotted by their percentage difference to experiment
($\delta\mathrm{BE} = (\mathrm{BE} - \mathrm{BE}^{\mathrm{exp}}) /
\mathrm{BE}^{\mathrm{exp}}$ and $\delta R_{\mathrm{ch}} =
(R_{\mathrm{ch}} - R_{\mathrm{ch}}^{\mathrm{exp}}) /
R_{\mathrm{ch}}^{\mathrm{exp}}$) and arranged by their mass number,
$A$, in \cref{suppfig1}. The experimental binding energies were taken
from Ref.~\cite{wang2017}, and the experimental charge radii were
taken from Ref.~\cite{angeli2013}. Because the CCSD approximation
omits triples correlations, the corresponding binding energies are
consistently underbound. The missing contribution from $3p$--$3h$
excitations can be approximated by adding $10\%$ of the correlation
energy to the CCSD results according to Refs.~\cite{bartlett2007,
  hagen2009b} which would reduce the underbinding
significantly. However, while binding energies may differ from
experiment by $> 10\%$, charge radii are consistently within a few
percent of experiment in the CCSD approximation.

\subsection{Additional results for binding energies and charge radii without the Coulomb force}

To isolate isospin-symmetry-breaking components of the nuclear
interaction, we compute binding energies and radii of select mirror
nuclei with $22\le A\le 56$ using interactions with and without the
Coulomb force. A subset of these results are shown in Table~I. The
remaining results, including the point-neutron and point-proton radii,
are shown below in \cref{tab:NoCoul1} and \cref{tab:NoCoul2} for
$\Delta$N$^2$LO$_{\rm GO}(394)$ and $\Delta$N$^2$LO$_{\rm GO}(450)$
interactions, respectively. Removing the Coulomb interaction has the
direct effect of contracting the proton distribution which results in
a contracted neutron distribution in turn. Because the first effect is
more pronounced, the neutron skin thickness slightly increases.

\begin{figure}[bth]
  \setlength\abovecaptionskip{-0.2\baselineskip}
  \hspace*{-8pt}
  \includegraphics[width=0.5\textwidth]{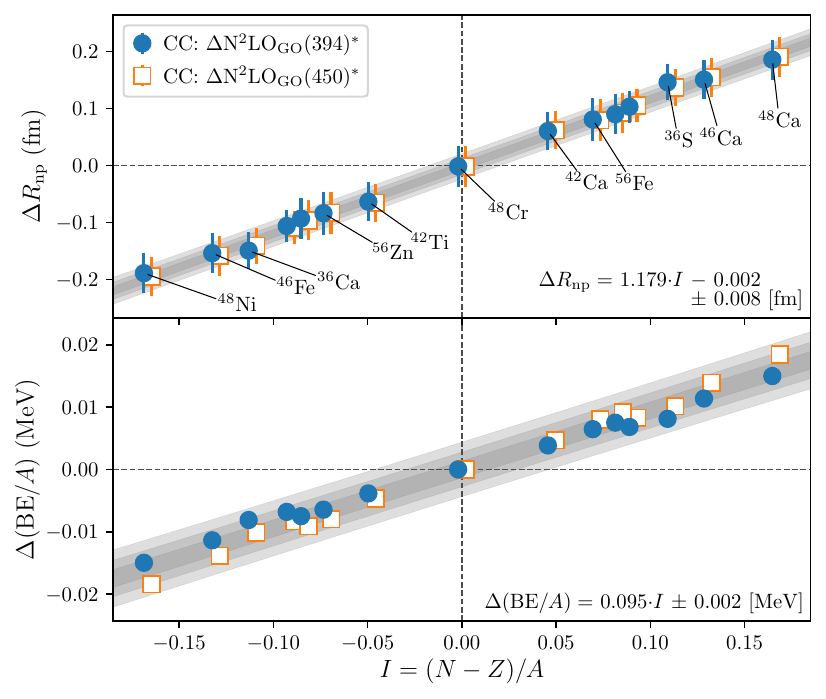}
  \caption{Top panel: Neutron skin thickness plotted against the
    isospin asymmetry from coupled-cluster calculations using the
    $\Delta$N$^2$LO$_{\rm GO}$(394) and $\Delta$N$^2$LO$_{\rm
      GO}$(450) interactions without the Coulomb force, shown as solid
    blue circles and empty orange squares, respectively. The linear
    regression is printed in the bottom right with $1\sigma$
    uncertainty, and the $1\sigma, 2\sigma$, and $3\sigma$ confidence
    levels are shown as grey bands. Select nuclei are labeled. Bottom
    panel: Mirror-difference binding energies per nucleon plotted
    against the isospin asymmetry from the same data in the top
    panel. The large error bars are omitted but detailed in
    \cref{tab:NoCoul1,tab:NoCoul2}. The linear regression is also
    shown similarly.}
  \label{suppfig2}
\end{figure}

The mirror-difference binding energies per nucleon and neutron skin
thicknesses computed without the Coulomb interaction are also shown in
\cref{suppfig2}, plotted against the isospin asymmetry ($I =
(N-Z)/A$). Both quantities exhibit linear relationships with the
isospin asymmetry. The uncertainties for the mirror-differences are
computed by adding the individual quantity uncertainties in quadrature
instead of taking the convergence error of the mirror-difference
itself, which tends to vanish and underestimate the
uncertainty. Therefore, the overestimated uncertainties of the
mirror-difference binding energies per nucleon are omitted from
\cref{suppfig2} but included in \cref{tab:NoCoul1,tab:NoCoul2}.

\subsection{Derivation of leading order corrections from the liquid-drop model}

In order to derive the simple trends that our \textit{ab initio}
results are expected to follow, we employs the liquid-drop model
(LDM)~\cite{Myers:1969, Myers:1974, Myers:1977, Myers:1980} that
describes nuclei in terms of their bulk properties. According to the
LDM, the binding energy of a nucleus can be characterized by its bulk
nuclear density ($\rho$), neutron density ($\rho_{n}$), and proton
density ($\rho_{p}$), where $\rho = \rho_{n} + \rho_{p}$. These
quantities can be used to construct variables that quantify small
deviations from $\rho=\rho_{0}$ and $\rho_{n} = \rho_{p}$,
\begin{equation}
  \delta = \frac{\rho_{n} - \rho_{p}}{\rho} \hspace{20pt}
  \varepsilon = -\frac{1}{3}\frac{\rho - \rho_{0}}{\rho_{0}},
\end{equation}
where $\rho_{0} = (4\pi r_{0}^{3}/3)^{-1}$ is the nuclear matter
density and $r_{0} = 1.126\ \mathrm{fm}$ is the nuclear radius
constant.

The LDM binding energy can be approximated~\cite{Myers:1969} by,
\begin{align}
  \setlength\abovedisplayskip{-5pt}
  \mathrm{BE} &= \left[ -a_{1} + J\delta^{2} + \frac{K}{2}
    \varepsilon^{2} \right] A \notag \\
  &+ \left[ a_{2} + \frac{4Q}{9} \left( I - \delta \right)^{2} A^{2/3}
    \right] A^{2/3} \notag \\
  &+ \frac{c_{1}}{2}\left[ Z\left( 1 - I \right)\left( 1 - \varepsilon -
    \frac{\delta}{3}\right) \right] A^{2/3}.
\end{align}
The first line consists of the volume terms of the energy, where
$a_{1}$ is the volume energy coefficient, $J$ is the symmetry energy
coefficient, and $K$ is the compressibility coefficient. The second
line consists of the surface terms, where $a_{2}$ is the surface
energy coefficient, $Q$ is the surface stiffness coefficient, and $I =
(N-Z)/A$ is the isospin asymmetry. The last line consists of the
Coulomb terms, where $c_{1} = 3e^{2}/5r_{0} \approx
0.767\ \mathrm{MeV}$ and the number of protons can be written as
$Z=A(1 - I)/2$. From Ref.~\cite{Myers:1969}, we use the coefficients
$J = 28.026\ \mathrm{MeV}$ and $Q = 16.04\ \mathrm{MeV}$; the other
coefficients either do not enter into the radii formula or are
extracted from data.

By minimizing the energy with respect to $\delta$ and $\varepsilon$,
and retaining leading orders in $I \rightarrow 0$ and $A\rightarrow
\infty$, the equilibrium parameters can be obtained as,
\begin{align}
  \bar{\delta} &= \left( I + \frac{3}{16} \frac{c_{1}}{Q} \left( 1 - I
  \right) Z A^{-1/3} \right) \left( 1 + \frac{9}{4}\frac{J}{Q}
  A^{-1/3} \right)^{-1}, \notag \\
  \bar{\varepsilon} &= \frac{c_{1}}{2K} \left( 1 - I \right) Z
  A^{-1/3}.
\end{align}
Next, the densities $\rho$, $\rho_{n}$, and $\rho_{p}$ are defined
by particle numbers and their respective radii,
\begin{align}
  \frac{4}{3}\pi R^{3}\rho &=  N + Z, \notag \\
  \frac{4}{3}\pi R_{\mathrm{n}}^{3}\rho_{n} &= N, \notag \\
  \frac{4}{3}\pi R_{\mathrm{p}}^{3}\rho_{p} &= Z.
\end{align}
Using these definitions, and expanding $(1 \pm I)^{1/3} \approx (1 \pm
I/3)$, the equilibrium neutron and proton radii can be written as,
\begin{align}
  R_{\mathrm{n}} &= r_{0}\left(2N\right)^{1/3}\left(1 +
  \bar{\varepsilon} - \frac{\bar{\delta}}{3}\right) \notag \\
  &\approx r_{0}A^{1/3}\left(1 + \frac{I}{3}\right)\left(1 +
  \bar{\varepsilon} - \frac{\bar{\delta}}{3}\right), \notag \\
  R_{\mathrm{p}} &= r_{0}\left(2Z\right)^{1/3}\left(1 +
  \bar{\varepsilon} + \frac{\bar{\delta}}{3}\right) \notag \\
  &\approx r_{0}A^{1/3}\left(1 - \frac{I}{3}\right)\left(1 +
  \bar{\varepsilon} + \frac{\bar{\delta}}{3}\right).
\end{align}

Now it is possible to obtain the leading order contributions from the
LDM for $\Delta R_{\mathrm{np}}({}^{A}_{Z}\mathrm{X}_{N}) =
R_{\mathrm{n}}({}^{A}_{Z}\mathrm{X}_{N}) -
R_{\mathrm{p}}({}^{A}_{Z}\mathrm{X}_{N})$ and $\Delta
R^{\mathrm{mirror}}_{\mathrm{np}}({}^{A}_{Z}\mathrm{X}_{N}) =
R_{\mathrm{n}}({}^{A}_{N}\mathrm{Y}_{Z}) -
R_{\mathrm{p}}({}^{A}_{Z}\mathrm{X}_{N})$. By again keeping the
leading orders, these can be written as,
\begin{align}
  \Delta R_{\mathrm{np}} &= \frac{2}{3}r_{0}A^{1/3}\left[\ I \left( 1
    + \bar{\varepsilon}\right) - \bar{\delta}\ \right] \notag \\
  &\approx r_{0}\left[ \frac{3}{2}\frac{J}{Q^{*}} I -
    \frac{1}{8}\frac{c_{1}}{Q^{*}} Z A^{-1/3} \right], \label{eq:Rnp}
\end{align}
\begin{align}
  \Delta R^{\mathrm{mirror}}_{\mathrm{np}} &= r_{0} \left( 2Z
  \right)^{1/3} \left[ -\bar{\varepsilon} +
    \bar{\varepsilon}\strut^{\hspace{0.75pt}
      \mathrm{mirror}} \hspace{-9pt} - \frac{ \bar{\delta} +
      \bar{\delta}\strut^{\mathrm{mirror}}}{3} \right] \notag \\ 
  &\approx r_{0}c_{1}\left[ \frac{1}{K} I A -
    \frac{1}{8}\frac{c_{1}}{Q^{*}} Z A^{-1/3} \right]. \label{eq:Rnpmirror}
\end{align}
Here, \eq{eq:Rnp} can be found in Refs.~\cite{Myers:1969, Myers:1974,
  Myers:1977, Myers:1980} and \eq{eq:Rnpmirror} was derived for this
work. The quantity $Q^{*}$ is the effective surface stiffness and
compressibility coefficient. This effective label accounts for the
non-negligible denominator in $\bar{\delta}$. It is defined as,
\begin{equation}
  Q^{*} \equiv Q \left( 1 + \frac{9}{4}\frac{J}{Q} A^{-1/3}
  \right),
\end{equation}
and can be approximated by setting the mass number to the maximum
value in this work, $A=56$, and using the ratio from
Ref.~\cite{Myers:1969} of $J/Q = 1.75$. Then, $Q^{*}$ takes on the
value $Q^{*} \approx 2.03\cdot Q = 32.5\ \mathrm{MeV}$.

\begin{table*}[!ht]
  \caption{Neutron-point radius ($R_{\mathrm{n}}$), proton-point
    radius ($R_{\mathrm{p}}$), neutron skin thickness ($\Delta
    R_{\mathrm{np}}$), and mirror-difference binding energy per
    nucleon ($\Delta ({\rm BE}/A)$) of select nuclei using
    $\Delta$N$^2$LO$_{\rm GO}(394)$ with and without the Coulomb
    term.}\label{tab:NoCoul1}
  \begin{ruledtabular}
    \begin{tabular}{c c c c c c c c c}
      & \multicolumn{2}{c}{$\underline{R_{\mathrm{n}}\ \rm (fm)}$} & \multicolumn{2}{c}{$\underline{R_{\mathrm{p}}\ \rm (fm)}$}
      & \multicolumn{2}{c}{$\underline{\Delta R_{\mathrm{np}}\ \rm (fm)}$} & \multicolumn{2}{c}{$\underline{\Delta ({\rm BE}/A)\ ({\rm MeV})}$} \\
      & w/ Coul. & w/o Coul. & w/ Coul. & w/o Coul. & w/ Coul. & w/o Coul. & w/ Coul. & w/o Coul. \\
      \hline & \\ [-2ex]
      $^{22}\rm Ne$ &  2.884(21)  &  2.863(20)  &  2.804(20)  &  2.760(20)  &  0.080(29)  &  0.103(28)  &  0.407(06)  &  0.007(05)  \\
      $^{36}\rm S$  &  3.248(24)  &  3.228(23)  &  3.140(22)  &  3.082(22)  &  0.108(32)  &  0.146(32)  &  0.718(13)  &  0.008(11)  \\
      $^{42}\rm Ca$ &  3.379(25)  &  3.352(24)  &  3.360(24)  &  3.292(23)  &  0.019(35)  &  0.060(34)  &  0.349(18)  &  0.004(15)  \\
      $^{46}\rm Ca$ &  3.480(26)  &  3.454(26)  &  3.366(24)  &  3.304(24)  &  0.113(35)  &  0.151(35)  &  1.039(25)  &  0.011(20)  \\
      $^{48}\rm Ca$ &  3.512(26)  &  3.489(26)  &  3.362(24)  &  3.303(23)  &  0.150(36)  &  0.186(35)  &  1.382(28)  &  0.015(23)  \\
      $^{48}\rm Ti$ &  3.514(26)  &  3.480(26)  &  3.465(26)  &  3.390(24)  &  0.049(37)  &  0.090(36)  &  0.689(29)  &  0.007(24)  \\
      $^{48}\rm Cr$ &  3.503(26)  &  3.461(25)  &  3.552(27)  &  3.463(25)  & -0.048(38)  & -0.001(36)  &  0.000(30)  &  0.000(26)  \\
      $^{56}\rm Fe$ &  3.650(28)  &  3.613(27)  &  3.616(27)  &  3.532(26)  &  0.034(39)  &  0.081(37)  &  0.665(42)  &  0.006(35)  \\
    \end{tabular}
  \end{ruledtabular}
  \caption{Neutron-point radius ($R_{\mathrm{n}}$), proton-point
    radius ($R_{\mathrm{p}}$), neutron skin thickness ($\Delta
    R_{\mathrm{np}}$), and mirror-difference binding energy per
    nucleon ($\Delta ({\rm BE}/A)$) of select nuclei using
    $\Delta$N$^2$LO$_{\rm GO}(450)$ with and without the Coulomb
    term.}\label{tab:NoCoul2}
  \begin{ruledtabular}
    \begin{tabular}{c c c c c c c c c}
      & \multicolumn{2}{c}{$\underline{R_{\mathrm{n}}\ \rm (fm)}$} & \multicolumn{2}{c}{$\underline{R_{\mathrm{p}}\ \rm (fm)}$}
      & \multicolumn{2}{c}{$\underline{\Delta R_{\mathrm{np}}\ \rm (fm)}$} & \multicolumn{2}{c}{$\underline{\Delta ({\rm BE}/A)\ ({\rm MeV})}$} \\
      & w/ Coul. & w/o Coul. & w/ Coul. & w/o Coul. & w/ Coul. & w/o Coul. & w/ Coul. & w/o Coul. \\
      \hline & \\ [-2ex]
      $^{22}\rm Ne$ &  2.920(21)  &  2.900(21)  &  2.838(20)  &  2.794(20)  &  0.082(29)  &  0.105(29)  &  0.407(045)  &  0.008(046)  \\
      $^{36}\rm S$  &  3.265(23)  &  3.246(23)  &  3.165(22)  &  3.109(22)  &  0.100(32)  &  0.137(32)  &  0.724(070)  &  0.010(072)  \\
      $^{42}\rm Ca$ &  3.395(24)  &  3.370(24)  &  3.374(24)  &  3.308(23)  &  0.021(34)  &  0.062(33)  &  0.349(081)  &  0.005(082)  \\
      $^{46}\rm Ca$ &  3.503(25)  &  3.479(25)  &  3.386(24)  &  3.324(24)  &  0.117(35)  &  0.155(34)  &  1.042(095)  &  0.014(095)  \\
      $^{48}\rm Ca$ &  3.538(26)  &  3.515(25)  &  3.383(24)  &  3.325(24)  &  0.155(35)  &  0.191(35)  &  1.387(102)  &  0.018(101)  \\
      $^{48}\rm Ti$ &  3.546(26)  &  3.513(25)  &  3.495(25)  &  3.420(24)  &  0.051(36)  &  0.093(35)  &  0.690(106)  &  0.009(105)  \\
      $^{48}\rm Cr$ &  3.534(26)  &  3.493(25)  &  3.583(27)  &  3.495(25)  & -0.049(37)  & -0.002(35)  &  0.000(110)  &  0.000(109)  \\
      $^{56}\rm Fe$ &  3.683(27)  &  3.648(27)  &  3.651(27)  &  3.569(25)  &  0.032(38)  &  0.079(37)  &  0.699(144)  &  0.008(132)  \\
    \end{tabular}
  \end{ruledtabular}
\end{table*}

\end{document}